\documentclass[fleqn,usenatbib]{mnras}
\pdfoutput=1

\usepackage[T1]{fontenc}
\usepackage{ae,aecompl}
\usepackage{array}

\usepackage{graphicx}	
\usepackage{amsmath}	
\usepackage{amssymb}	
\usepackage{longtable}

\title[NS-WD mergers]{Neutron star - white dwarf mergers:\\
Early evolution, physical properties, and outcomes}

\author[Zenati, Perets \& Toonen]{
Yossef Zenati$^{1}$, 
Hagai B. Perets$^{1}$ 
and Silvia Toonen$^{1,2}$
\\
$^{1}$Physics Department, Technion - Israel Institute of Technology,
Haifa 3200004, Israel\\
$^{2}$Anton Pannekoek Institute for Astronomy, University of Amsterdam,
1090 GE Amsterdam, The Netherlands
}

\date{Accepted XXX. Received YYY; in original form ZZZ}
 
\pubyear{2018}

\hypersetup{draft}

\begin{document}
\label{firstpage}
\pagerange{\pageref{firstpage}--\pageref{lastpage}}
\maketitle

\begin{abstract}
Neutron-star (NS) - white-dwarf (WD) mergers may give rise to observable
explosive transients, but have been little explored. We use 2D coupled hydrodynamical-thermonuclear
FLASH-code simulations to study the evolution of WD debris-disks formed
following WD-disruptions by NSs. We use a 19-elements nuclear-network
and a detailed equation-of-state to follow the evolution, complemented
by a post-process analysis using a larger 125-isotopes nuclear-network.
We consider a wide range of initial conditions and study the dependence
of the results on the NS/WD masses ($1.4-2{\rm M_{\odot}}$;$\,{\rm 0.375-0.7\,M_{\odot}}$, respectively), WD-composition (CO/He/hybrid-He-CO) and the accretion-disk
structure. We find that viscous inflow in the disk gives rise to continuous
wind-outflow of mostly C/O material mixed with nuclear-burning products
arising from a weak detonation occurring in the inner-region of the
disk. We find that such transients are energetically weak ($10^{48}-10^{49}$ergs)
compared with thermonuclear-supernovae (SNe), and are dominated by
the (gravitational) accretion-energy. Although thermonuclear-detonations
occur robustly in all of our simulations (besides the He-WD) they
produce only little energy $(1-10\%$ of the kinetic energy) and $^{56}{\rm Ni}$
ejecta (few$\times10^{-4}-10^{-3}{\rm M_{\odot}})$, with overall low ejecta masses of $\sim0.01-0.1{\rm M_{\odot}}$. Such explosions may produce rapidly-evolving transients, much shorter and fainter than regular type-Ia SNe.
The composition and demographics of such SNe appear to be inconsistent
with those of Ca-rich type Ib SNe. Though they might be related to
the various classes of rapidly evolving SNe observed in recent years,
they are likely to be fainter than the typical ones, and may therefore
give rise a different class of potentially observable transients.
\end{abstract}




The physical outcomes and observable expectations from mergers of
neutron-stars (NSs) and white-dwarfs (WDs) are not well understood,
and have been relatively little explored. \citet{1999ApJ...520..650F}
and \citet{Kin+07} suggested that accretion of the white-dwarf debris
on a neutron star may produce a unique type of a long GRB. More recently,
Metzger and collaborators \citep{2012MNRAS.419..827M,2013ApJ...763..108F,Mar+16}
studied the early accretion phase of the WD-debris and the evolution
of the accretion disk. They proposed that thermonuclear reactions
can play an important role in the evolution of the disk, even prior
to the final accretion of material on the NS, and suggested that such
nuclear dominated accretion can give rise to faint thermonuclear explosion
occurring in the accretion disk \citep[see also][]{Mar+16,Mar+17}.
Here we follow these studies and explore NS-WD mergers and the evolution
of the WD-debris disk and the accretion driven outflows through the
use of more detailed and realistic models, that alleviate many of
the potential difficulties and uncertainties in the previous models,
as we describe below.

\citet{2013ApJ...763..108F} employed 2D (axisymmetric) hydrodynamical
simulations of radiatively inefficient accretion flows with nuclear
burning. They studied the vertical dynamics of the disc and its interplay
with the radially steady burning front. They found that the nuclear
energy released at the burning front could be larger than the local
thermal energy. When this condition is satisfied the burning front
can spontaneously transition into an outwards propagating detonation,
due to the mixing of hot downstream matter (ash) with cold upstream
gas (fuel). Such detonations either falter once the shock propagates
into the outer regions of the disc, or completely disrupt the large-scale
accretion flow. However, \citet{2013ApJ...763..108F} noted that the
detonations they observe could be an artifact of their simplified
equation of state (EOS), which included only gas pressure and neglected
radiation pressure (thus artificially accentuating the temperature
discontinuity at the burning front). \citet{2013ApJ...763..108F}
also employed only a single nuclear reaction, which prevented them
from making detailed predictions for the composition of the disc outflows
and their electromagnetic signatures. This also required them to add
an ad-hoc parameter as to achieve more efficient nuclear burning,
required to ensue a detonation. Finally, they neglected the self-gravity
of the disk, which could change the disk structure and evolution.

Similar to the approach introduced by \citet{2013ApJ...763..108F},
we follow the evolution of an accretion-disk formed following the
disruption of a WD by a NS, using simple, but physically motivated
assumptions for the initial structure of the disk. We use the publicly
available FLASH v4.2 code \citep{2000ApJS..131..273F} to generate
2D hydro simulations of the disk, but we improve on the previous modeling
in various aspects. In terms of the physical modeling we include a
detailed 19-elements nuclear reaction network; a more realistic Helmholtz
EOS and we account for the self-gravity of the disk. These allow us
to adequately and self-consistently capture the nucleosynthetic energetics
without any ad-hoc assumptions. In addition we follow-up the simulations
with detailed post-process analysis of the nucleosynthetic products
from the disk using an extended $125$ isotopes network.

Besides the more sophisticated models and the nucleosynthetic post-processing
analysis, we also explore a wide range of initial conditions. We vary
the NS and WD masses, as well as the WD composition (in particular
considering hybrid - He-CO WDs), and we consider a range of initial
accretion disk configurations. Together these allow us to study the
dependence of the outcomes of the disk evolution on a wide range of
parameters. We generally find that the accretion-disk evolution robustly
gives rise to weak thermonuclear explosions, but these produce little $^{56}{\rm Ni}$
(at most $10^{-3}$ ${\rm M_{\odot}})$ which, by themselves, could
only give rise to very faint transients. Nevertheless, the accretion
process releases much more significant energy; even a fraction of
which could potentially give rise to a more energetic transient if
converted to electromagnetic emission. The dependence of the outcomes
on the initial configurations of such mergers is described in details.
Finally, 2D detailed simulations are too computationally expensive
as to allow for long-term evolution studies as done in the simplified
1D disk simulations of \citet{Mar+16}, however we study a test-case
of a lower-resolution long-term (30 s) simulation.

We begin by describing our simulations and the various types of initial
conditions we explored in Section 2, we then describe the main results
in Section 3, and discuss and summarize them in Section 4.

\section{Methods and initial conditions}

The evolution of a disrupted WD debris disk around a NS is simulated
using the publicly available FLASH v4.2 code \citep{2000ApJS..131..273F}.
The simulations were done using the unsplit ${\rm PPM}$ solver of
FLASH in ${\rm 2D}$ axisymmetric cylindrical coordinates on a grid
of size ${\rm 1\times1\left[10^{10}cm\right]}$ using adaptive mesh
refinement. We follow similar approaches as described in other works
on thermonuclear SNe (e.g. \citealt{Mea+09}). Detonations are handled
by the reactive hydrodynamics solver in FLASH without the need for
a front tracker, which is possible since unresolved Chapman--Jouguet
(CJ) detonations retain the correct jump conditions and propagation
speeds. Numerical stability is maintained by preventing nuclear burning
within the shock. This is necessary because shocks are artificially
spread out over a few zones by the ${\rm PPM}$ hydrodynamics solver,
which can lead to nonphysical burning within shocks that can destabilise
the burning front \citep{1989BAAS...21.1209F}. In addition we consider
a wider range of initial conditions for the structure, mass and composition
of the disk, and we consider two different masses for the accreting
NS.

We use a detailed EOS and account for the self-gravity of the disk.
We also employ a $19$-isotopes reaction network, which burning front
\citep{1989BAAS...21.1209F}. In order to prevent the production of
artificial unrealistic early detonation that may arise from insufficient
numerical resolution, we applied a limiter approach following \citet{2013ApJ...778L..37K}.

We made multiple simulations with increased resolution until convergence
was reached in the nuclear burning. We found a resolution of $1-10$
km to be sufficient for convergence of up to 10$\%$ in energy. Gravity
was included as a multipole expansion of up to multipole $l=12$ using
the new FLASH multipole solver, to which we added a point-mass gravitational
potential to account for gravity of the NS. We simulated the viscus
term by using the the viscosity unit in Flash, employing a \citet{1973A&A....24..337S}
parameterization ${\rm \nu_{\alpha}=\alpha C_{s}^{2}/\Omega_{Kepler}}$,
where ${\rm \Omega_{Keplere}}$ is the Keplerian frequency and ${\rm C_{s}}$
is the sound speed,  and the $\alpha$ parameter used is $0.01$. The contributions of both nuclear reaction and
neutrino cooling \citep{1989ApJ...346..847C,1991ApJ...376..234H}
are included in the the internal energy calculations, and the Navier-Stocks
equations are solved with source terms due to gravity, shear viscosity
and the nuclear reactions.

The equation of state (EOS) used in our simulations is the detailed
Helmoholz EOS employed in FLASH (\citealt{2000ApJS..126..501T}).
This EOS includes contributions from partially degenerate electrons
and positrons, radiation, and non-degenerate ions. It uses a look-up
table scheme for high performance. The most important aspect of the
Helmholtz EOS is its ability to handle thermodynamic states where
radiation dominates, and under conditions of very high pressure.

The nuclear network used is the FLASH $\alpha-$chain network of 19
isotopes. This network can adequately capture the energy generated
during the nuclear burning \citep{2000ApJS..126..501T}. In order
to follow the post-process analysis of the detailed nucleosynthetic
processes and yields we made use of $4000-10000$ tracer particles
that track the radius, velocity, density, and temperature and are
evenly spaced every ${\rm 2\times10^{8}cm}$ throughout the WD-debris
disk. Our simulations are evolved for $7-13$ seconds. In one test
case we run a lower resolution evolution up to 30 seconds. Following
the FLASH runs we make use of the detailed histories of the tracer
particles density and temperature to be post-processed with MESA (version
8118) one zone burner \citep{2015ApJS..220...15P}. We employ a 125-isotope
network that includes neutrons (see supplementary information), and composite reactions
from JINA's REACLIB \citep{2010ApJS..189..240C}. Overall we find
that the results from the larger network employed in the post-process
analysis show less efficient nuclear burning (it is a negligible), and giving rise to somewhat
higher yields of intermediate elements on the expense of lower yields
of iron elements, similar to the results seen in other works \citep{2013MNRAS.436.3413G,2016ApJ...822...19P}.

\subsection{Initial disk properties}

We focus on disks that form when a WD is tidally disrupted by a companion
NS in a close binary system \citep{1999ApJ...520..650F}. We first
review the characteristic properties of the disks, closely following
\citet{2013ApJ...763..108F}. Whether the WD is disrupted by the NS
depends on stability of the the mass-transfer process following the
onset of Roche lobe overflow. Once the WD is disrupted, conservation
of angular momentum implies that the material will circularize around
the NS at a characteristic radius 
\begin{equation}
{\rm R_{0}=\frac{a_{RLOF}}{1+q}}\label{eq:sim_r0}
\end{equation}
where ${\rm a_{RLOF}=f\cdot R_{WD}}\:,0.4<f<0.8$ , $R_{WD}$ is the
radius of the WD, and the mass ratio of the binary is given by ${\rm q=M_{WD}/M_{NS}}$.
The orbital time at the circularization radius is

\begin{equation}
{\rm t_{orb}}\simeq{\rm 38\left(\frac{R_{0}}{10^{9.3}\rm cm}\right)^{3/2}\left(\frac{M_{NS}}{1.4M_{\odot}}\right)^{-1/2}s}\label{eq:timescal orb}
\end{equation}
The characteristic timescale for matter to accrete is estimated by
the viscous time and the characteristic accretion ${\rm \dot{M}\sim M_{WD}/t_{visc}}$.
Following \citet{1999MNRAS.310.1002S} and \citet{2013ApJ...763..108F},
the torus density is normalized to its maximum value $\rho_{max}$
, thereby fixing the polytropic constant in terms of the adiabatic
index and the torus distortion parameter $d$. The latter is a measure
of the internal energy content of the torus and the scale-height $H_{0}$
\citep{1999MNRAS.310.1002S,2013ApJ...763..108F}. We then get

\begin{equation}
\rho_{{\rm disk}}=\rho_{{\rm max}}\left[\left(\frac{2{\rm H}}{{\rm R_{0}}}\right)\frac{2{\rm d}}{{\rm d-1}}\left(\frac{{\rm R_{0}}}{r}-\frac{1}{2}\left(\frac{{\rm R_{0}}}{{\rm r}\sin\theta}\right)^{2}-\frac{1}{2{\rm d}}\right)\right]^{7/2}\label{eq:rhodisk}
\end{equation}

\begin{equation}
{\rm \frac{P}{\rho}}={\rm \frac{2GM}{5R_{0}}}{\rm \left[\frac{R_{0}}{r}-\frac{1}{2}\left(\frac{R_{0}}{r\sin\theta}\right)^{2}-\frac{1}{2d}\right]}\label{eq:Press}
\end{equation}
Note we would improve on \citet{2013ApJ...763..108F} that have not accounted for the self-gravity of the disk.  \citet{2013ApJ...763..108F} used a simplified $\gamma=5/3$ EOS and did not solve for the value of d, and considered 3 somewhat arbitrary specific values. In contrast, we have self-consistently derived the structure of the disk, by first choosing initial values for the parameters in Eq. \ref{eq:rhodisk} like  \citet{2013ApJ...763..108F}, but then we derived the actual EOS given the disk conditions in FLASH. We then rederived the structure of the disk using the new EOS parameters. We iteratated this procedure until the structure converged, i.e. the EOS and the density distribution in each of the disk region cells in FLASH did nit change (to the level of ${10^{-5}}$ between consecutive iterations. We find that our more consistent model for disk structure, accounting for more realistic EOS and the self-gravity produce more compact disks than assumed by \citet{2013ApJ...763..108F}
  
We assume that the internal energy is dominated by non-degenerate
particles, and that it balances $25\%$ of the gravitational energy,
this assumption derives from the virial theorem. Note that the opacity
is dominated by electron scattering and the diffusion timescale for
photons escape the disk is much longer than the timescale for the
disk formation ${\rm t_{diff}\gg t_{orb},t_{visc}}$. Where ${\rm t}_{{\rm visc}}$
is the viscous timescale 
\begin{eqnarray}
\label{eq:tacc}
t_{\rm visc}& \simeq &\alpha^{-1}\left(\frac{R_{0}^{3}}{GM_{\rm c}}\right)^{1/2}\left(\frac{H_{0}}{R_{0}}\right)^{-2}\nonumber\\
&\sim& 2600{\rm\,s} \left(\frac{0.01}{\alpha}\right)\left(\frac{R_{0}}{10^{9.3}\rm cm}\right)^{3/2}
\left(\frac{1.4M_{\sun}}{M_{\rm c}}\right)^{1/2}\left(\frac{H_{0}}{0.5R_{0}}\right)^{-2}\nonumber\\
\end{eqnarray}
where $\alpha$ parametrizes the disk viscosity.

Neutrino cooling (included in our simulations) does not play an important
role. Even at the hottest and most dense regions we find that the
timescale for neutrino cooling is far longer than the simulation time.
We further validated this through running similar simulations without
neutrino cooling - the results show essentially no difference from
the simulation which did include this process.

\subsection{WD-debris disk and NS Models }

The detailed properties of each of the NS-WD models we explored are
described in table \ref{tab:WD-models}. The mass and composition
of the WD-debris models we consider are determined by the properties
of the WD progenitors of the disk. The properties of the WDs are obtained
through detailed stellar evolution models of single and binary stars
using the MESA code \citep{2011ApJS..192....3P,2015ApJS..220...15P}.
In all cases we considered only Solar metallicity stellar progenitors.
Our models include both typical CO WDs as well as hybrid HeCO WDs.
The former are produced from the regular evolution of single stars,
that eventually produce WDs composed of $\sim$50$\%$ carbon and
$\sim$50$\%$ oxygen. The hybrid WDs, containing both CO and He are
derived from detailed \emph{binary} evolution in MESA, as described
in \citet{Zen+18}. We also considered several artificial WD compositions
with higher He fractions than produced in our models. Though these
may potentially arise from accretion of He on a WD under some complex
binary evolutionary scenarios, we stress that these have significantly
less physical motivation from stellar evolution models, and should
be considered with a grain of salt.

The NSs are not resolved in our simulations and only participate in
the simulations as point masses/potentials. We considered two NS masses,
$1.4{\rm M_{\odot}}$and $2{\rm M_{\odot}}$. Together the NS mass
and the WD structure (mass and composition) determine the tidal radius
at which a given WD is expected to be disrupted ($r_{t}\sim{\rm (M_{NS}/M_{WD})^{1/3}R_{WD}}$;
where ${\rm M_{NS}},$ ${\rm M_{WD}}$ and ${\rm R_{WD}}$ are the
NS mass, WD mass, and WD radius, respectively). The disk outer-radius
is assumed to be positioned near the tidal radius, but we have explored
a range of specific inner disk-radii (see table ).

Simulations of WD disruptions by a stellar compact objects (e.g. \citealt{1998ApJ...502L...9F,2012MNRAS.422.2417D})
suggest very thick disks are produced, but the exact structure of
the disk is not known. We therefore considered two disk heights ${\rm H/R_{0}=0.5}$
and $0.7$. The initial structure of the disk is assumed to follow
a Shakura-Sunyaev structure, as described above. Note that we find
that the self-gravity of the disk, not included in previous studies,
can significantly alter the structure of the disk even before significant
radial evolution occurs in the disk. The effects of the self-gravity
become more pronounced for smaller disk heights.

\begin{table*}
\begin{centering}
\begin{tabular}{|c|c|c|c|c|c|c|c|c|c|}
\hline 
\# & ${\rm M_{WD}}$  & ${\rm M_{NS}}$  & ${\rm \rho_{max}[g]}$  & ${\rm R_{0}/r_{t}}$  & ${\rm \nu\,cooling}$  & ${\rm H/R_{0}}$  & ${\rm \%He_{4}}$  & ${\rm \%C_{12}}$  & ${\rm \%O_{16}}$\tabularnewline
\hline 
\hline 
A{*} & $0.53$  & $1.4$  & $1.9\times10^{6}$  & $1.1$  & Yes  & $0.5$  & $-$  & $50$  & $50$\tabularnewline
\hline 
B & $0.5$  & $1.4$  & $5.5\times10^{6}$  & $0.8$  & No  & $0.5$  & $-$  & $50$  & $50$\tabularnewline
\hline 
C & $0.55$  & $1.4$  & $8.5\times10^{6}$  & $0.8$  & No  & $0.5$  & $-$  & $50$  & $50$\tabularnewline
\hline 
D & $0.62$  & $1.4$  & $6.4\times10^{6}$  & $1$  & No  & $0.5$  & $9$  & $50$  & $41$\tabularnewline
\hline 
E{*}{*} & $0.62$  & $1.4$  & $8.5\times10^{6}$  & $0.8$  & No  & $0.5$  & $4$  & $49$  & $47$\tabularnewline
\hline 
F & $0.62$  & $2.01$  & $2.3\times10^{6}$  & $1.1$  & Yes  & $0.5$  & $-$  & $50$  & $50$\tabularnewline
\hline 
G & $0.73$  & $1.4$  & $2.1\times10^{7}$  & $1$  & No  & $0.5$  & $-$  & $50$  & $50$\tabularnewline
\hline 
H & $0.73$  & $2.01$  & $4.9\times10^{5}$  & $2$  & No  & $0.5$  & $-$  & $50$  & $50$\tabularnewline
\hline 
I & $0.73$  & $1.4$  & $1.1\times10^{7}$  & $1$  & No  & $0.7$  & $-$  & $50$  & $50$\tabularnewline
\hline 
J{*} & $0.8$  & $1.4$  & $4.4\times10^{7}$  & $0.8$  & No  & $0.5$  & $-$  & $50$  & $50$\tabularnewline
\hline 
K  & $0.28$  & $1.4$  & $5.2\times10^{4}$  & $1.1$  & No  & $0.5$  & $100\%$  & $-$  & $-$\tabularnewline
\hline 
L  & $0.28$  & $1.4$  & $1.3\times10^{4}$  & $2$  & No  & $0.5$  & $100\%$  & $-$  & $-$\tabularnewline
\hline 
\end{tabular}
\par\end{centering}
\caption{The parameters of the simulated NS-WD merger
models.  ${\rm M_{WD}}$  is the mass of the disrupted WD. ${\rm M_{NS}}$  is the mass of the NS. ${\rm \rho_{max}[g]}$ is the maximum density in the debris disk.  ${\rm R_{0}/r_t}$ is the distance of the innermost edge of the disk in units of the tidal-disruption radius. ${\rm \nu\,cooling}$ describes whether neutrino cooling was considered.  ${\rm H/R_{0}}$ is the height scale of the disk.   ${\rm \%He_{4}}$, ${\rm \%C_{12}}$ and ${\rm \%O_{16}}$ are the fractions of He, C and O composing the disk. \protect \\
{*}These models were run both with and without neutrino cooling, with
no differences observed\protect \\
{*}{*}These models were also run for longer time, but at a lower resolution.
 }
\label{tab:WD-models}
\end{table*}

\section{RESULTS}

Our main results are summarized in table \ref{tab:results} where
the overall properties of the simulated models are described. The
cases of He WDs did not give rise to any thermonuclear explosive event, nor to outflows; 
the disk expanded but no mass was ejected from the system, nor any nuclear
burning occurred on the timescales of our simulations. We therefore
do not further discuss these models which appear produce a peculiar
but non-explosive object; the long-term evolution of such an objects
is worth exploring, but is beyond the scope of this paper . In the
following we discuss the evolution of the other models and their outcomes
in more detail.

\begin{figure*}
\includegraphics[scale=0.2]{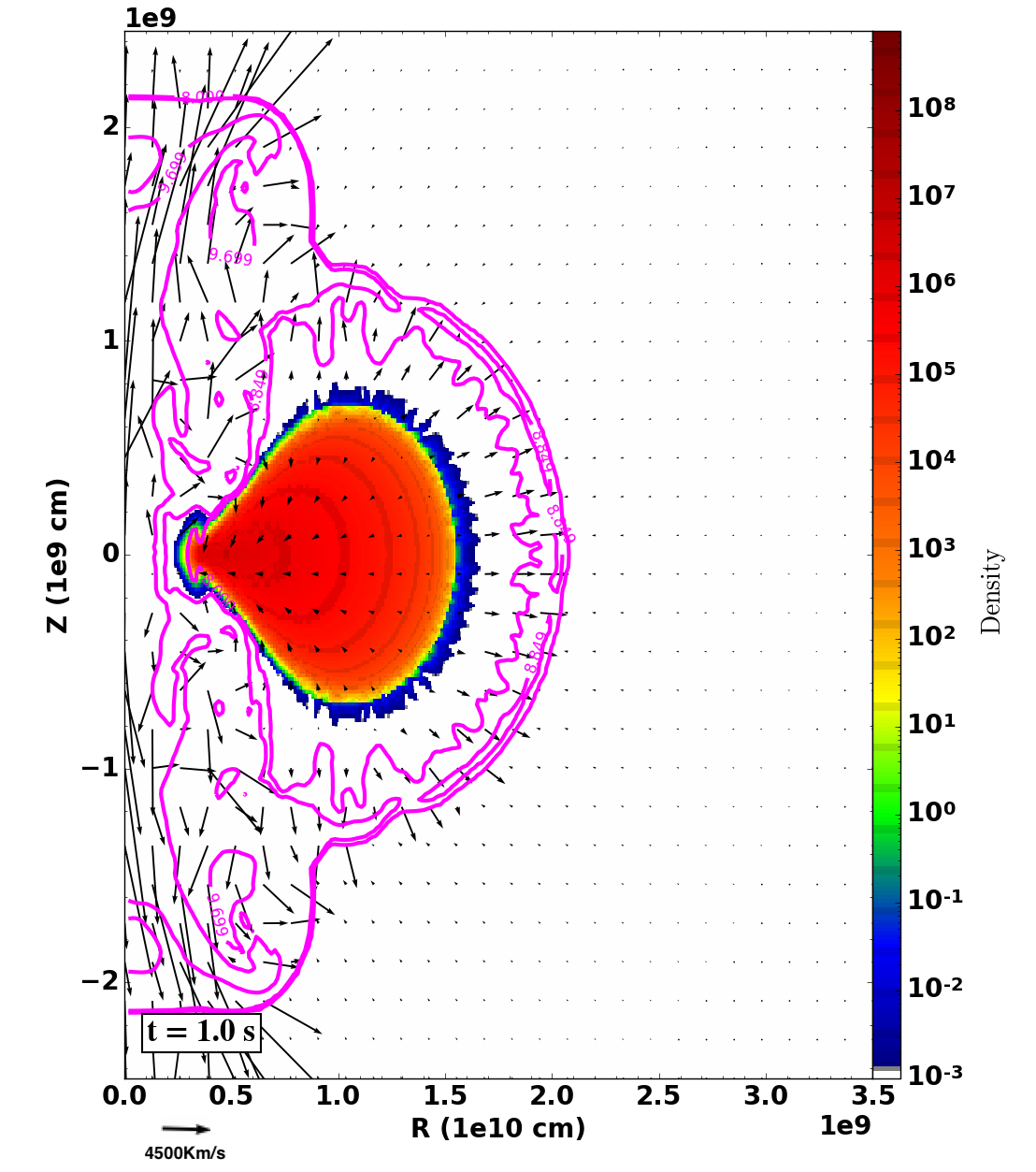}\includegraphics[scale=0.2]{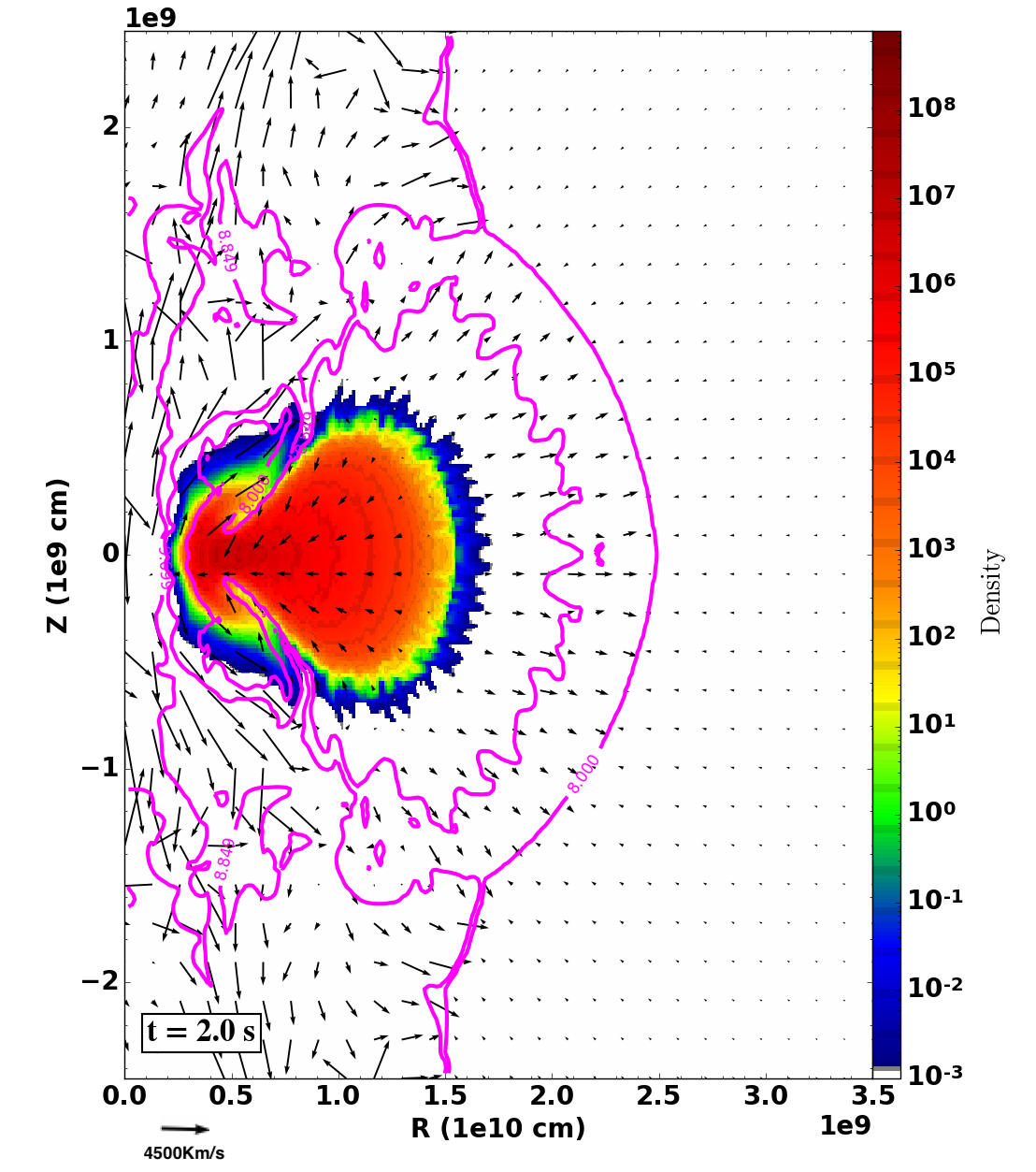}\includegraphics[scale=0.2]{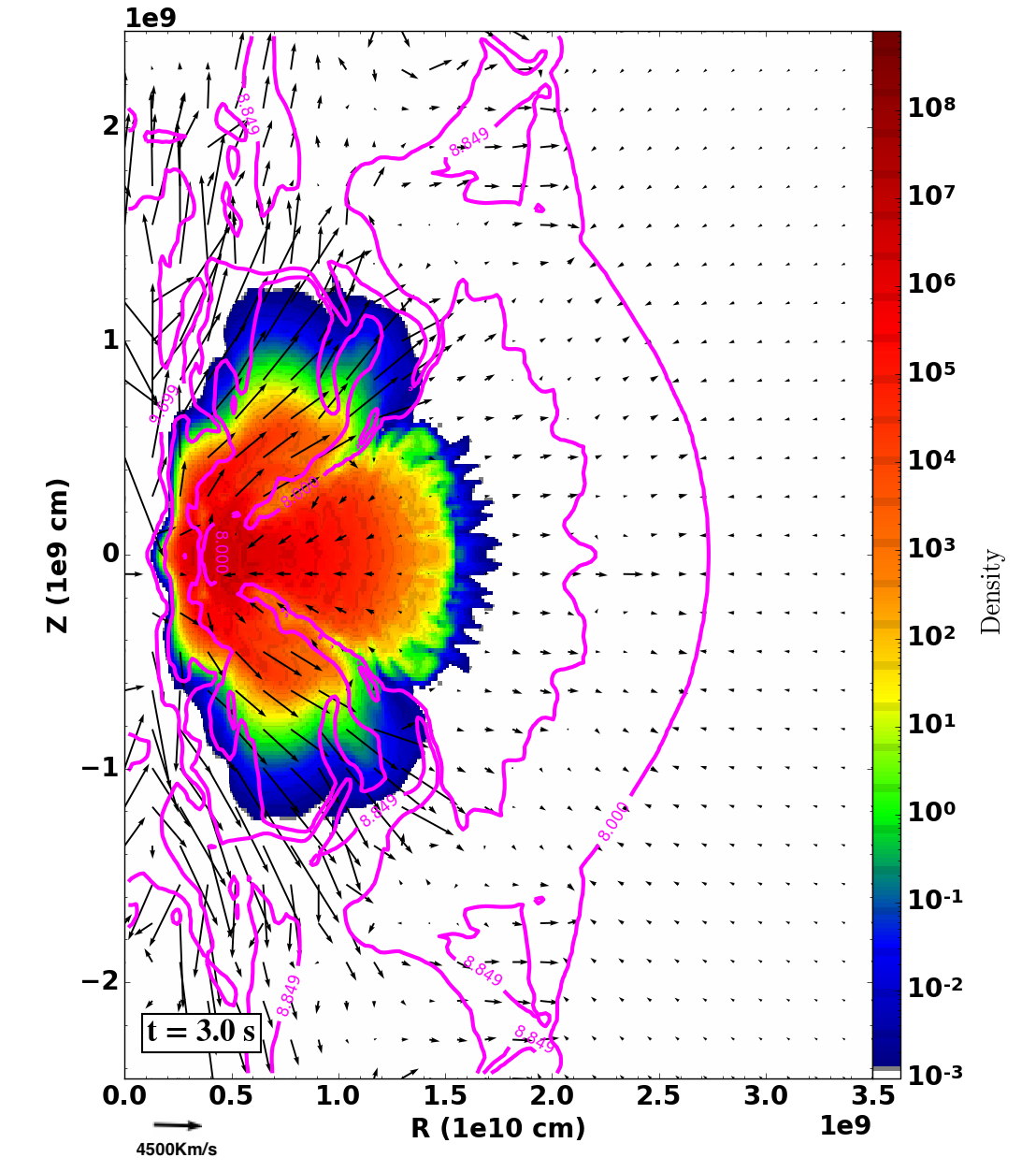}

\includegraphics[scale=0.2]{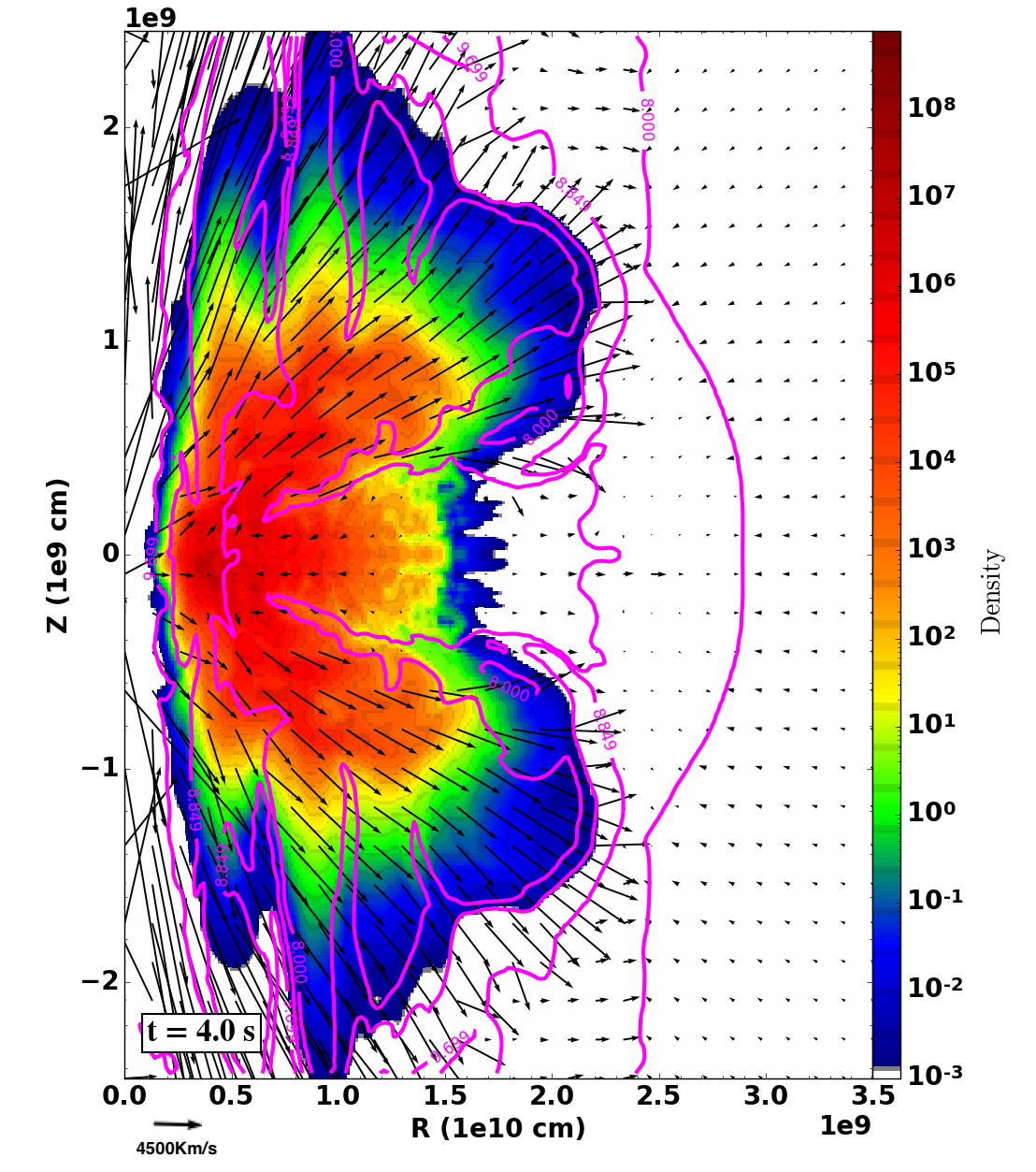}\includegraphics[scale=0.2]{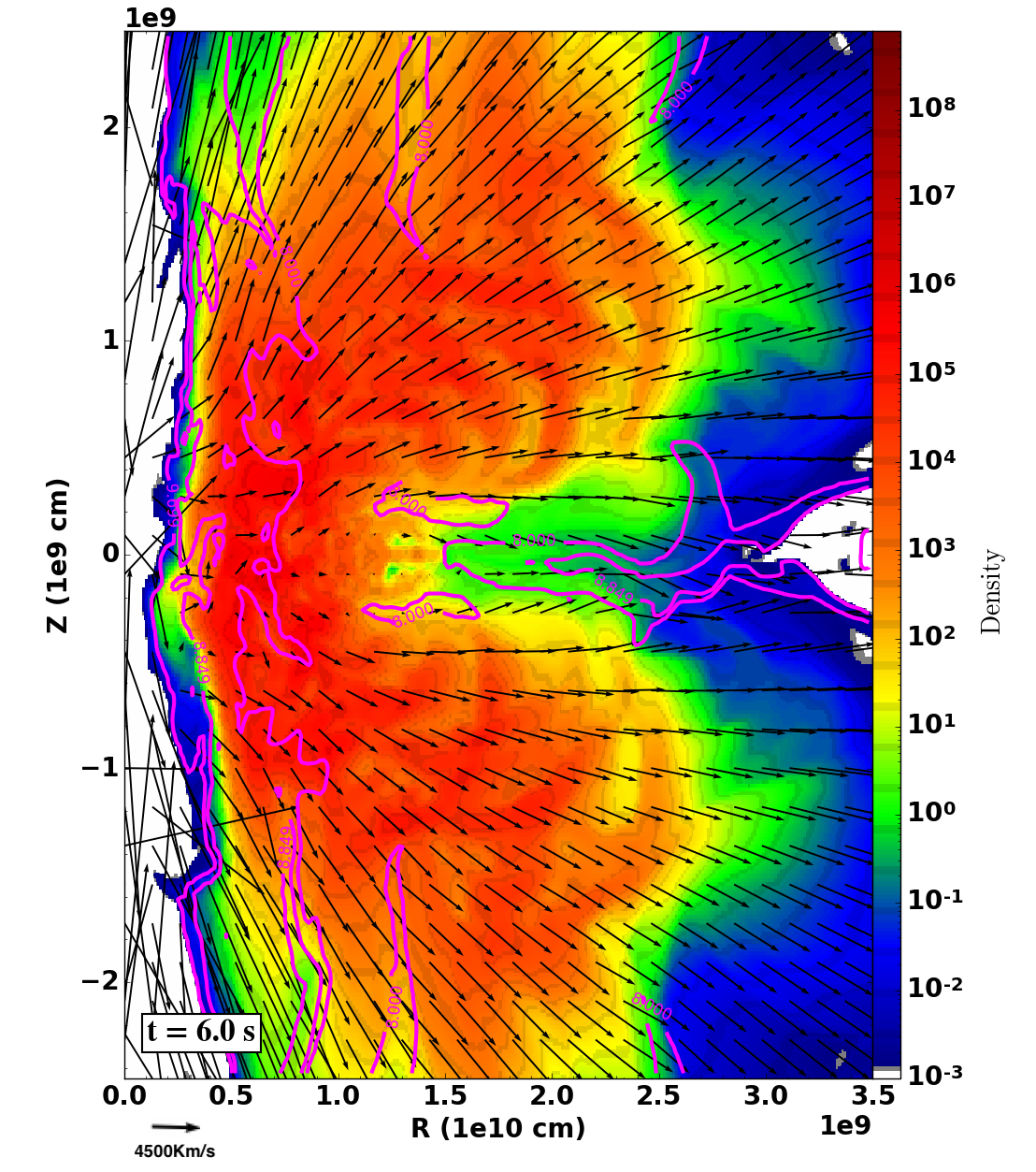}\includegraphics[scale=0.2]{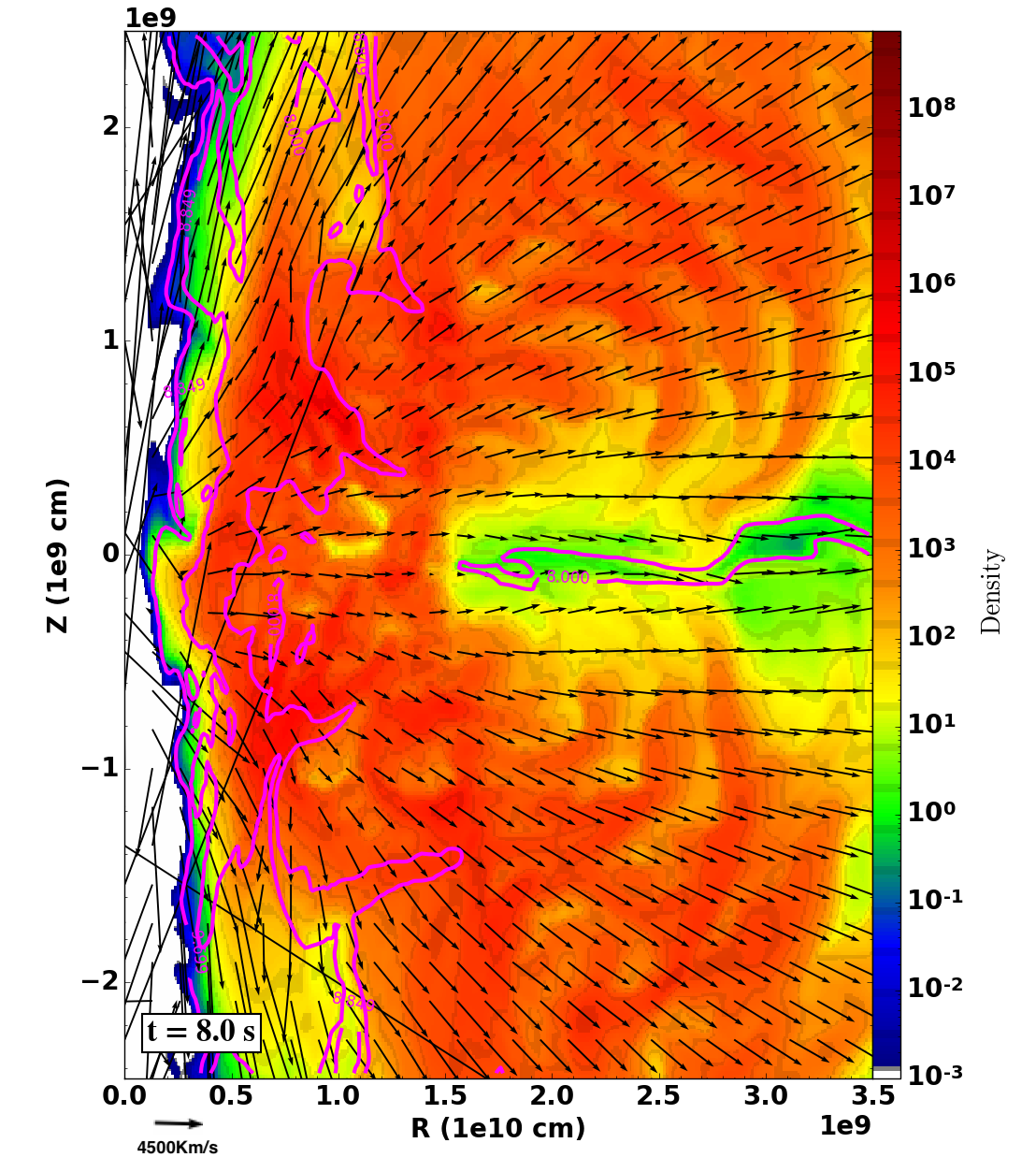}

\caption{The evolution of the WD debris
at early times for model E, modeled at high resolution. Each panel
shows the (color coded) density distribution and velocity vectors
(black arrows) at different time. The velocity scale is 4500 ${\rm km\cdot sec^{-1}}$.
Magenta contours correspond to the temperature.}
\label{fig:disk-evolution}
\end{figure*}

Though the disk evolution and final outcomes depend on the specific
model or initial conditions, the overall evolution of the disks we modeled
follow a very similar evolution, and we focus on one example shown
in Fig. \ref{fig:disk-evolution}. As can be seen in the snapshots
the disk evolves radially through viscous evolution; the inner regions
of the disk spread inwards, while the outer regions expand outwards.
At the same time the inner parts of the disk gravitationally collapse
vertically to form a thinner structure. As the inner disk evolves
into a more compact and thinner configuration its density and temperature
increase. Material then accretes inwards through the (horizontal)
central parts of the disk and outflows of material ensues after a
few seconds. The outflows are ejected at a wide angle from the innermost
central parts, as gravitational accretion energy is converted into
heat and kinetic energy fuelling the outflows. The fastest and hottest
material is funnelled almost vertically from the inner-most parts of
the accretion disk, producing a jet-like structure with velocities
extending up to $\sim3\times10^{4}$km s$^{-1}$, but enclosing only
a small fraction of the ejected material. 

Note that material is accreting inwards throughout the simulation, but the main accretion happens at early times, until heated material in the inner regions gives rise to significant pressure and outflows. These, in turn choke further significant accretion, which then happens only stochastically and only through low-rate infall of material. At these later times outflows from the inner region counteract most of the inflow in the disk producing a region where material is stalled at about $2.5\times10^{8}$ cm. Note that the inner regions showing apparently ``empty'' regions, but as shown by the velocity arrows are not empty but rather correspond to lower densities below the color-coding resolution.

As the density and temperature increase in the inner regions they
attain the critical conditions for thermonuclear burning, and an explosive
weak detonation occurs. However, only a small fraction of the accreting
material participates in the thermonuclear burning and therefore little
amounts of nucleosynthetic by-products are produced, with only up
to $\sim10^{-3}-10^{-2}$${\rm M_{\odot}}$ of material is burnt into
intermediate or even iron elements. In particular, at most $10^{-3}$
${\rm M_{\odot}}$ of $^{56}{\rm Ni}$ are produced, and only little
nuclear energy ($\sim{\rm few\,\times}10^{46}-10^{47}$ erg) is produced
during the process (see Table \ref{tab:results}). For comparison,
10-100 times larger gravitational energy $(\sim{\rm few\,\times}10^{47}-10^{49}\,{\rm erg})$
is released, driving the outflows and heating of the material. In
other words, nuclear processes appear to play a relatively minor role
in the overall evolution of the debris-disk evolution and outflows. In particular, a model run without
and nuclear interactions gave rise to very similar kinematic results
compared with its counterpart run which included nuclear burning,
with the main difference being a small fraction of ejected burnt material
in the latter run.

\begin{figure}
\includegraphics[scale=0.3]{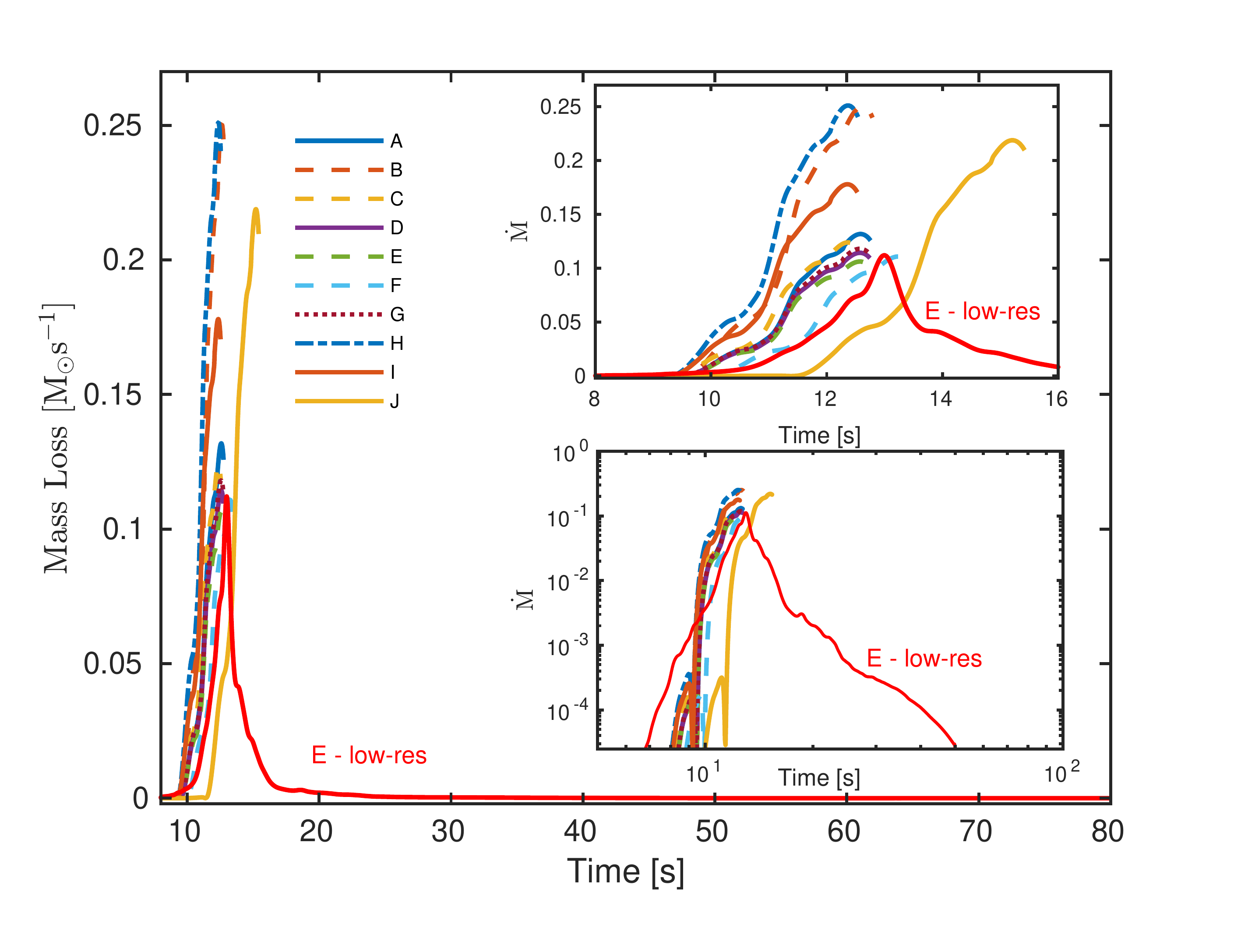}

\caption{The rate of mass-loss from the system (unbound
material) in each of the models as a function of time. The subplots
show a zoom-in of the same evolution at early times, and the same on logarithic scale. Both short-term high resolution and long-term low-resolution simulations are shown
for model E. }
\label{fig:mass-loss}
\end{figure}

The computational expense limits the length of our simulations. However
as a test case we run two of our simulations at lower resolution and
followed them on a larger simulation box, better allowing us to follow
the long term evolution of the outflows. As can be seen in Fig. \ref{fig:disk-evolution}
the early evolution followed both by the low and high resolution simulations
is qualitatively similar, suggesting that the low-resolution models
reasonably follow the NS-WD evolution. The long-term evolution show
that after the first 10 seconds or so outflows transform the initially
thin disk into a puffy structure around the NS with extended outflows
ejecting a few percents of the WD-debris. The long-term evolution
of the now more spherical cloudy structure of the leftover debris
around the NS is beyond the scope of our models, to be studied elsewhere.

We note that the comparison between the low-resolution long-term models
with the high-resolution short-term simulations suggest that the overall
energetics and mass-loss in the short-term models represent only about
half of the total energetics/outflows arising from the merger.

In Fig. \ref{fig:mass-loss} we show the mass-loss evolution for the
various models. It shows the results from the high resolution simulations
run up to $8-12$ s. For model E we also run a longer-term, but lower-resolution
simulation up to 80 seconds, which compares well with its low-resolution
counterpart simulation. As can be seen, the outflow rate increases
to peak after a few seconds and then decreases back to a negligible
level.

We analyze the rate of mass crossing the closest region near the
NS resolved in our simulations ($2\times10^{8}$ cm), in order to
provide an upper limit on the accretion rate onto the NS. We can not
resolve the size of the NS, however, even assuming that all the mass
that crosses into this region accretes on the NS, the total mass accreted
is small, not more than $10^{-4}$ ${\rm M_{\odot}}$, not likely
to produce a typical or even a sub-luminous GRB.  

\begin{figure*}
\includegraphics[scale=0.2]{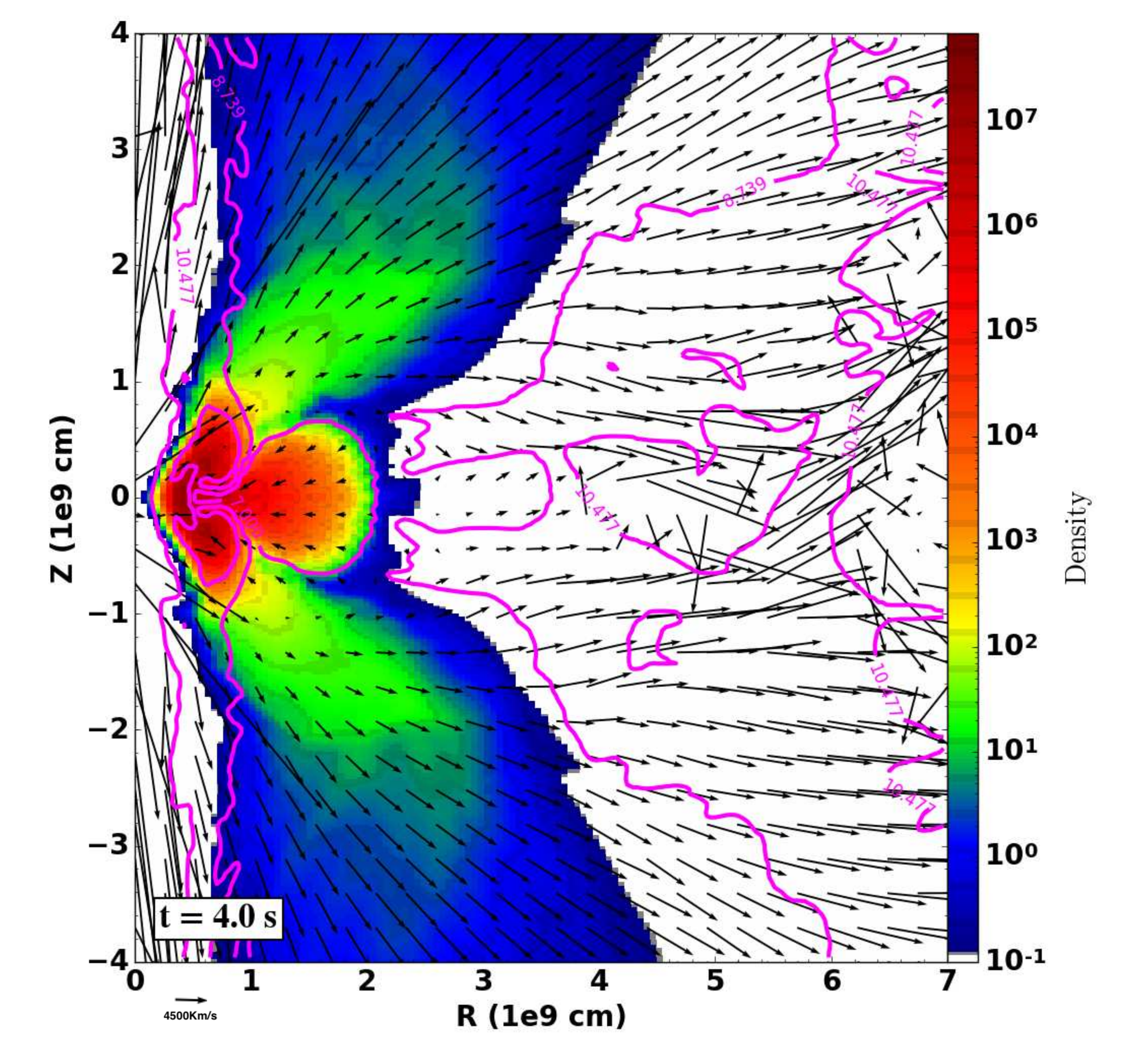}\includegraphics[scale=0.2]{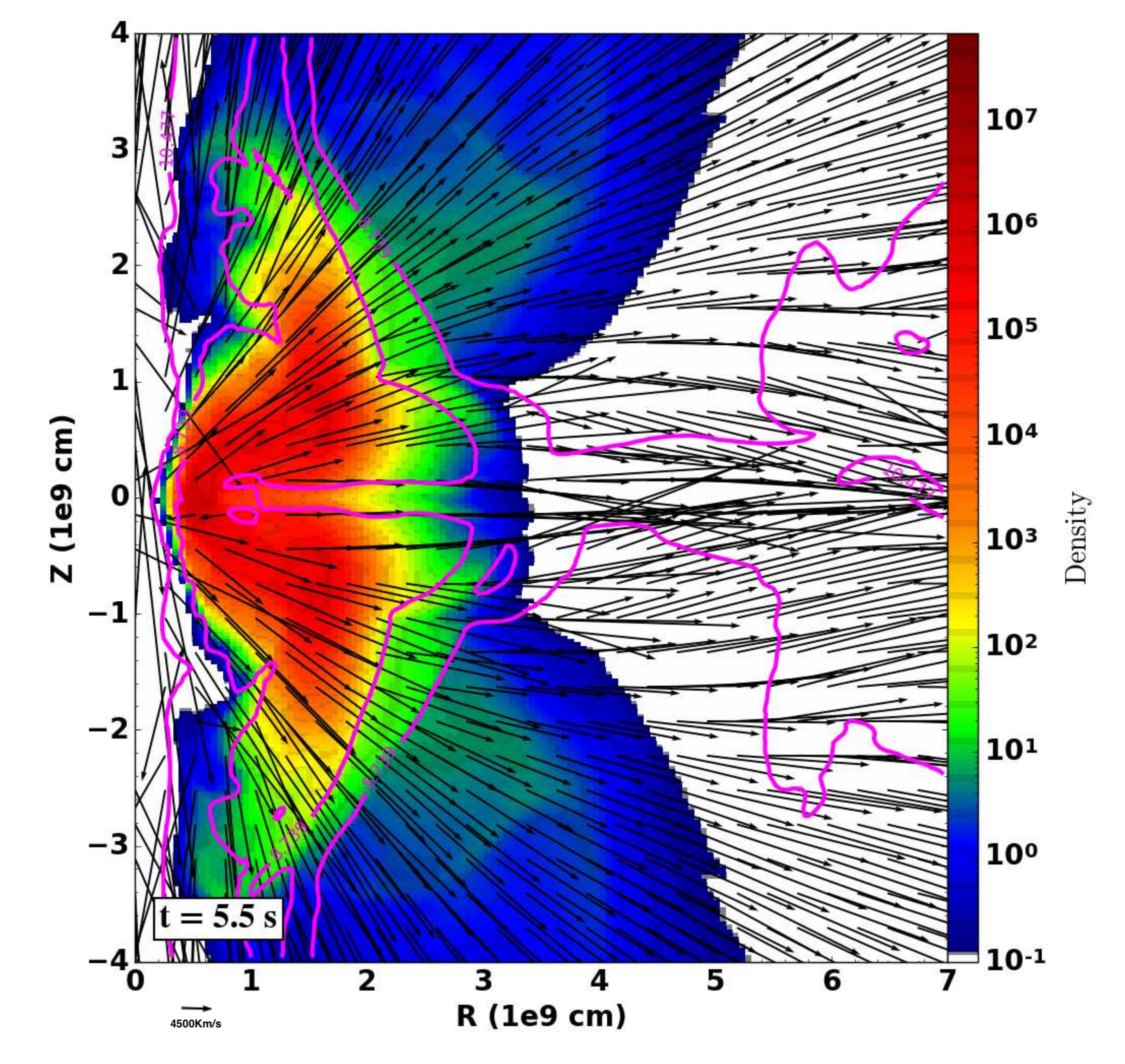}\includegraphics[scale=0.2]{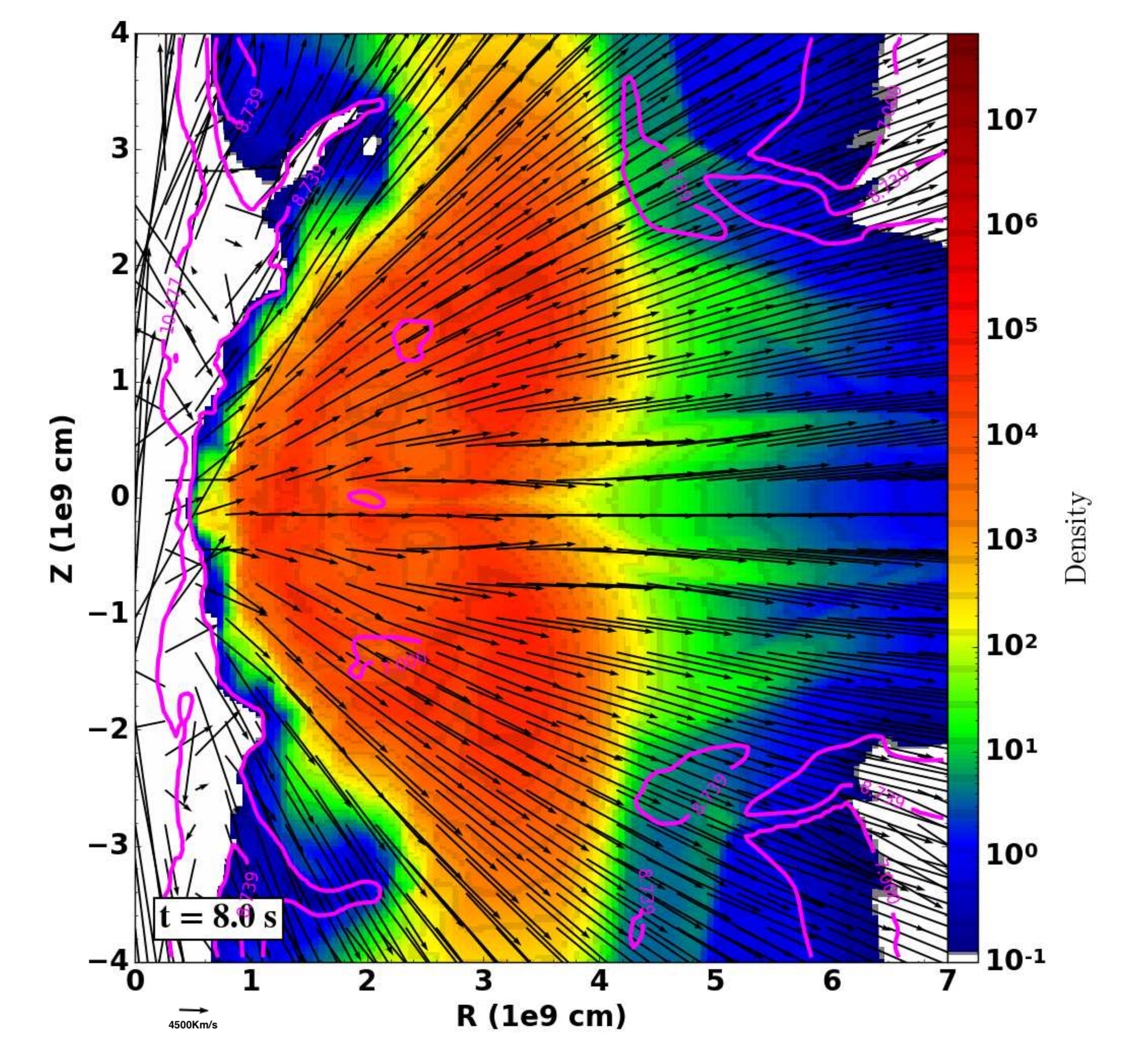}

\includegraphics[scale=0.2]{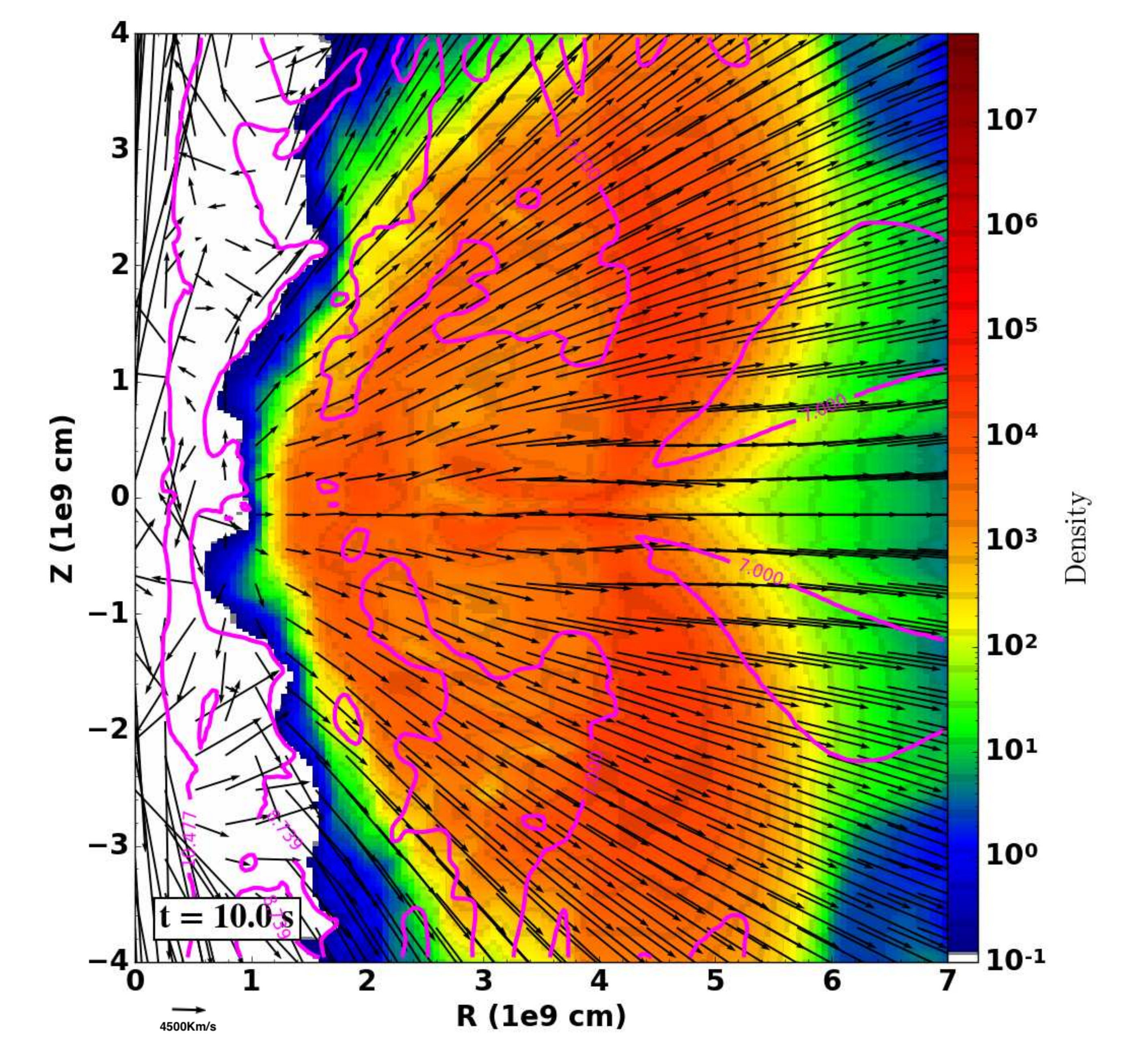}\includegraphics[scale=0.2]{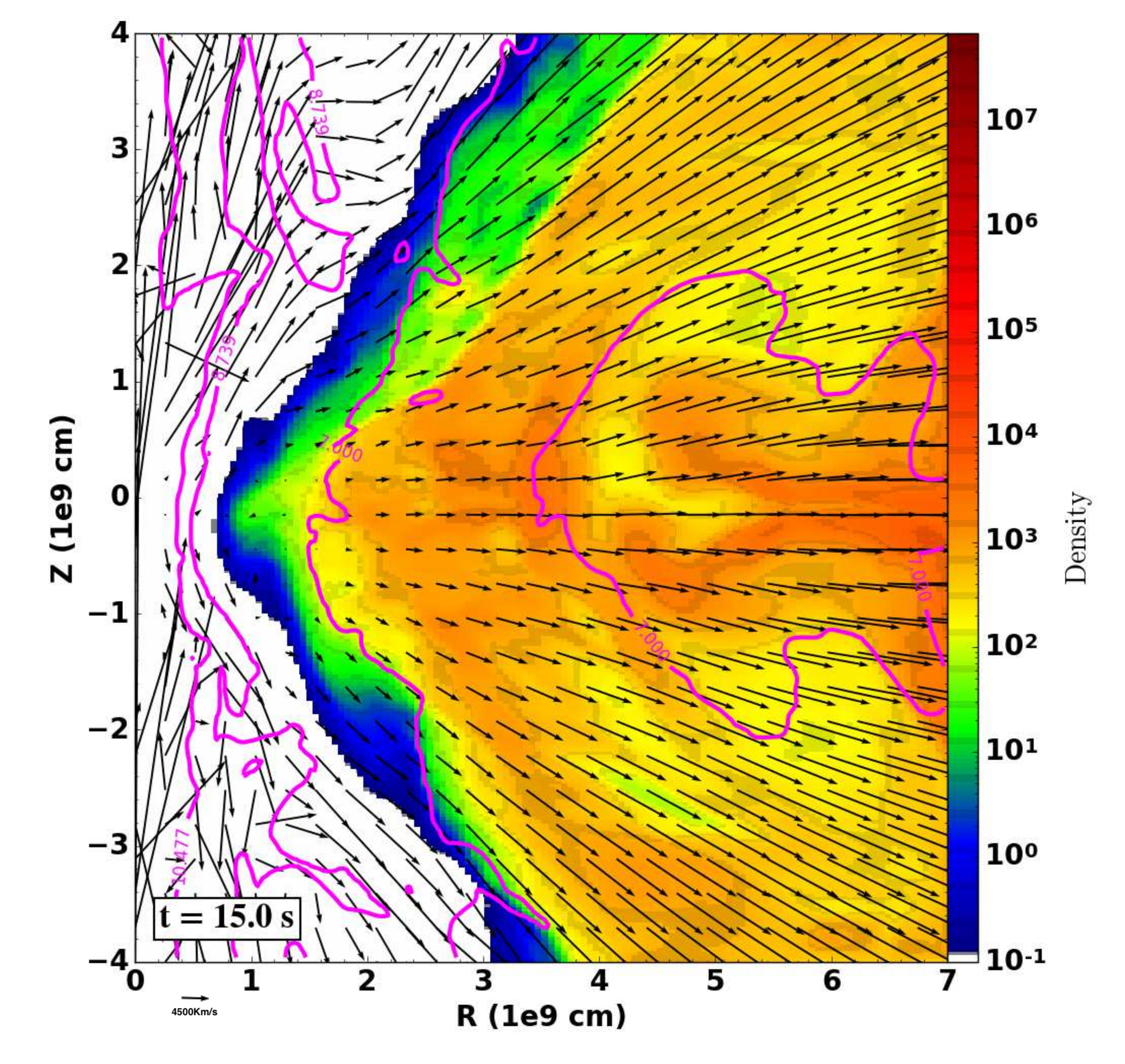}\includegraphics[scale=0.2]{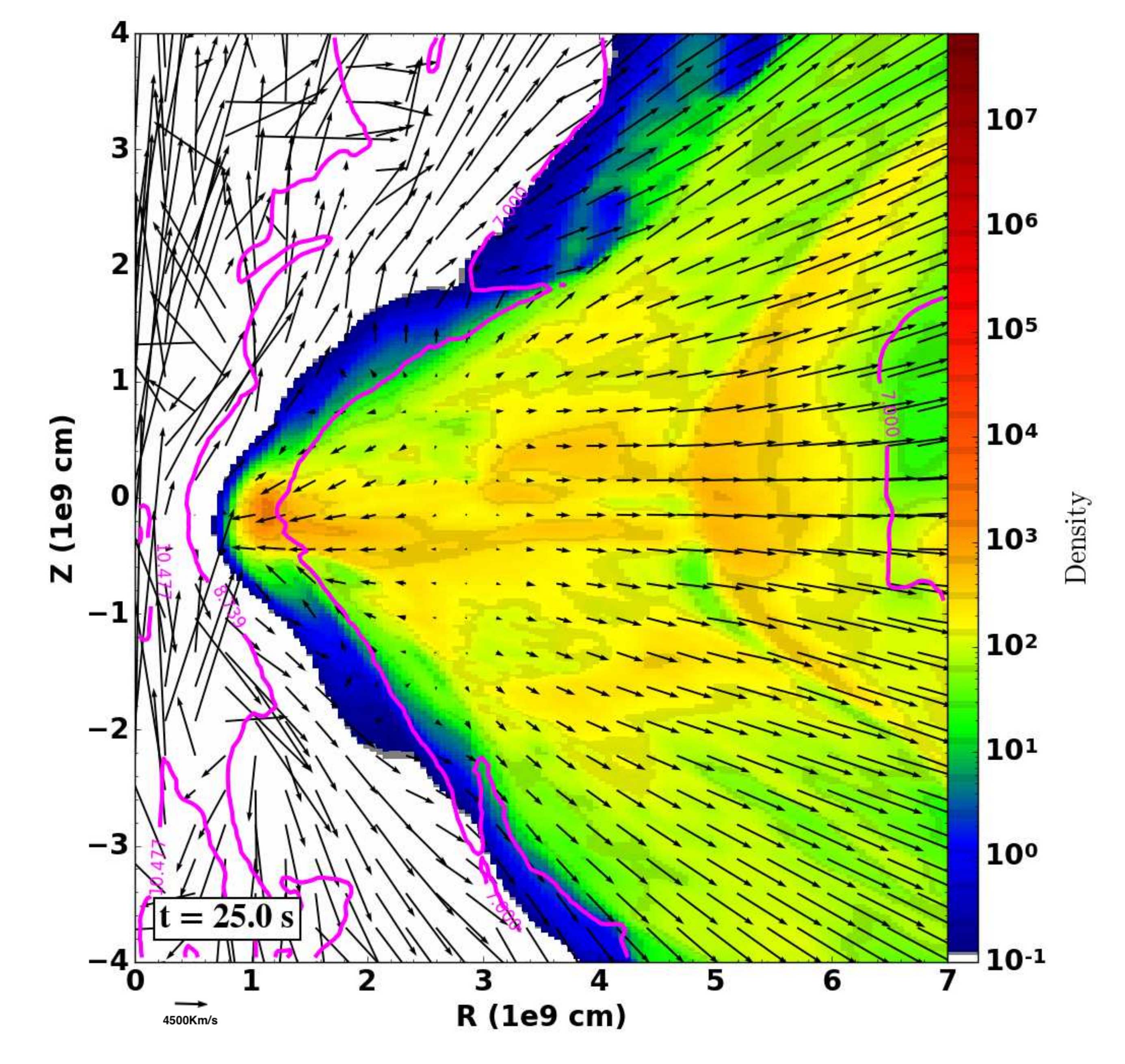}

\includegraphics[scale=0.2]{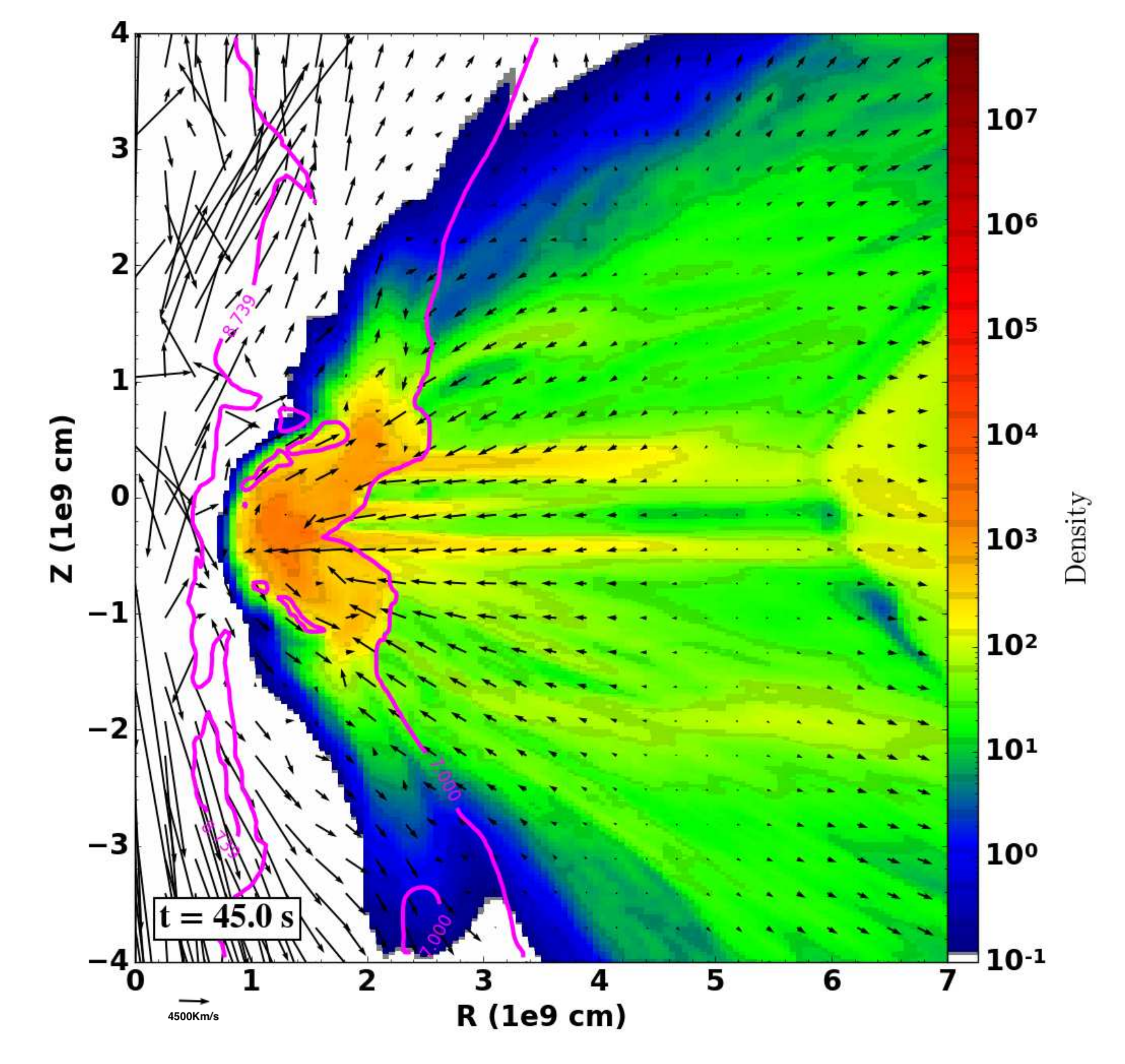}\includegraphics[scale=0.2]{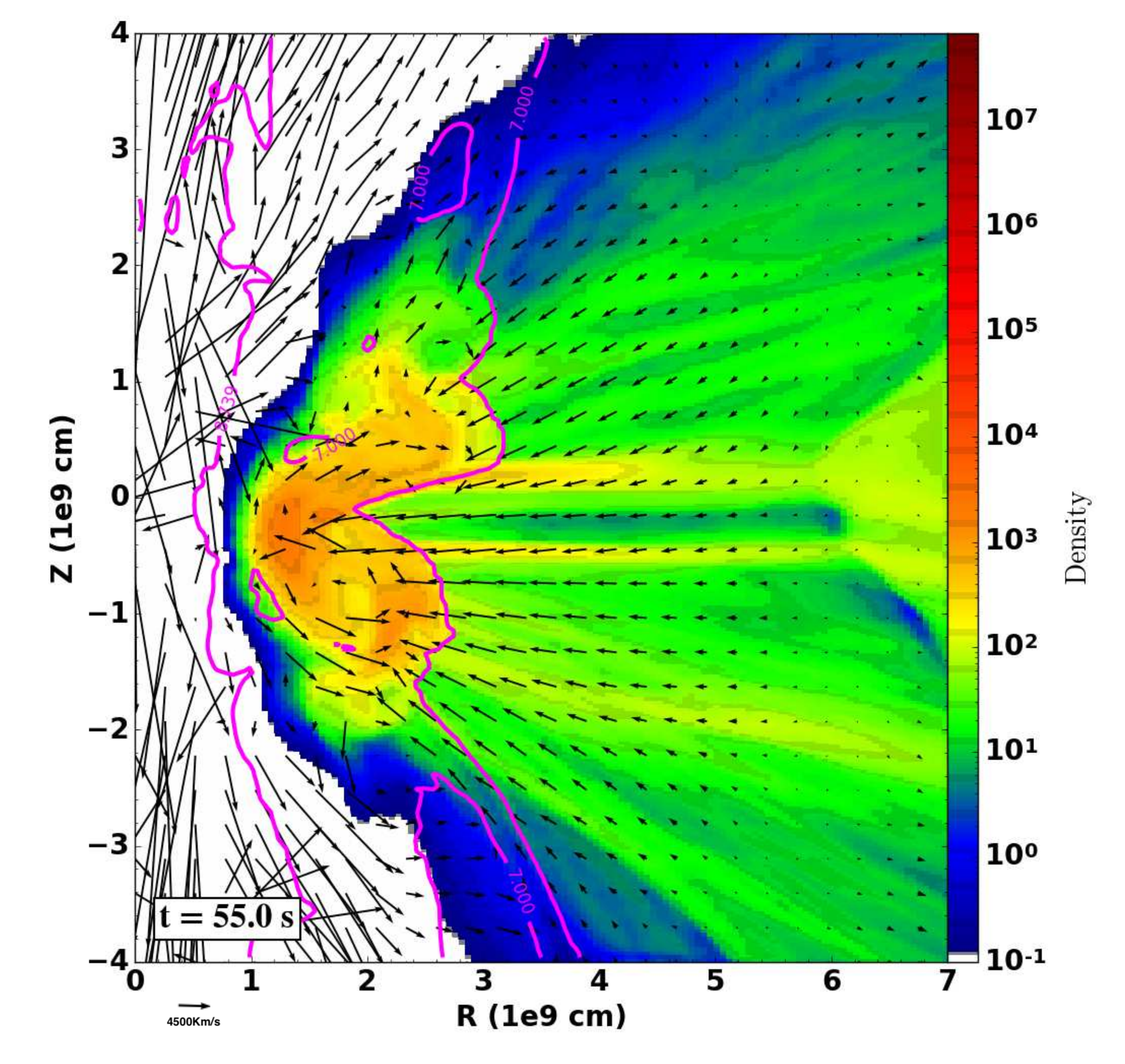}\includegraphics[scale=0.2]{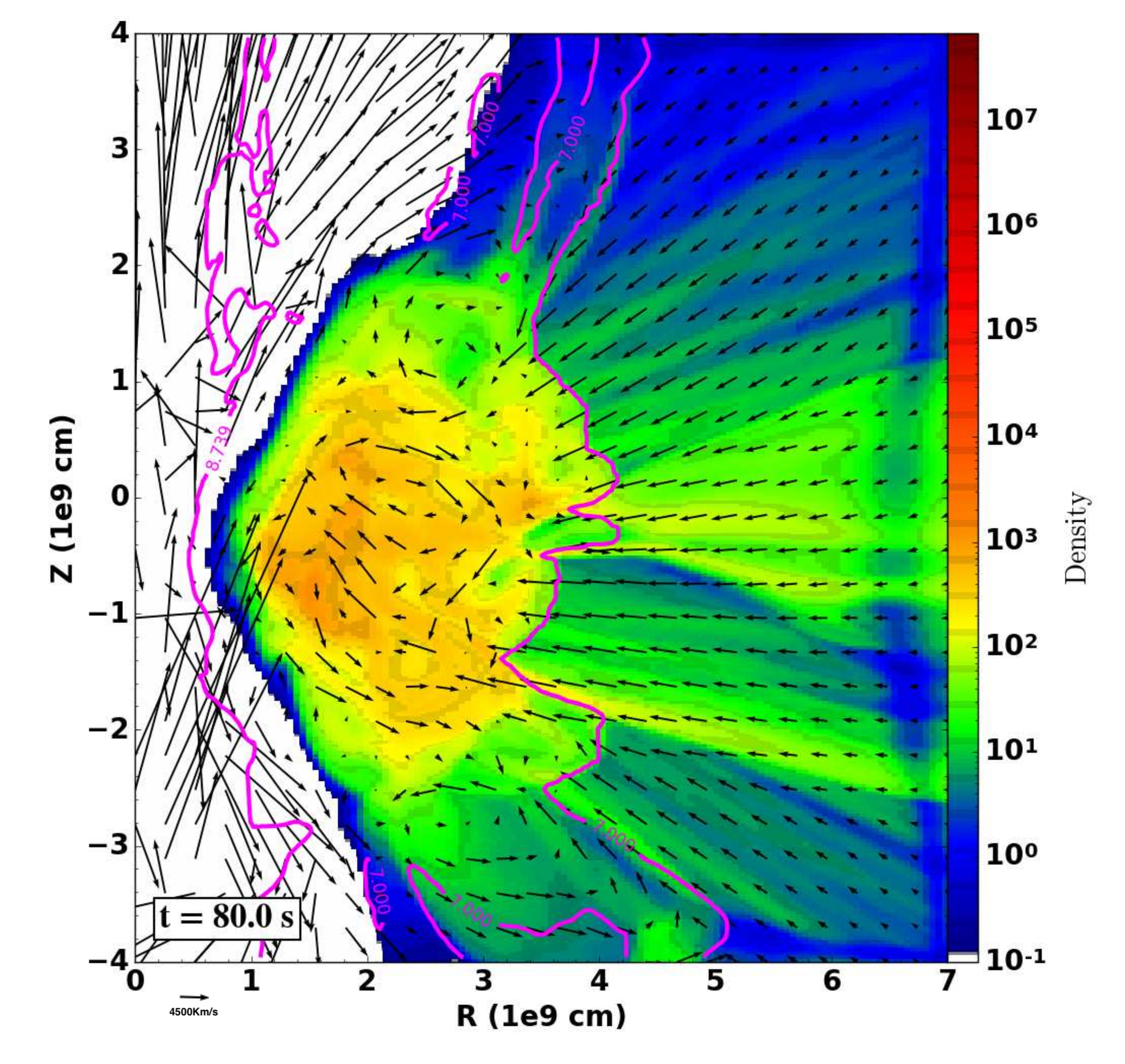}

\caption{\label{fig:disk evolution low res}The long-term evolution of the
WD debris outflows (similar to Fig. \ref{fig:disk-evolution})
in from the low-resolution simulation of model E; . There are three
different velocity scales 5000, 4600, and 3000 ${\rm km\cdot sec^{-1}}$
for three different epochs $0-5,5-20$ and $25-80{\rm \ sec}$, respectively. }
\end{figure*}

\begin{table*}
\caption{\label{tab:results}}
\begin{tabular}{|c|c|c|c|c|c|c|c|c|c|c|c|>{\centering}p{1cm}|}
\hline 
\# & {\scriptsize{}{}${\rm M_{WD}}$ }  & {\scriptsize{}{}${\rm E_{thermal}}$ }  & {\scriptsize{}{}${\rm E_{nuc}}$ }  & {\footnotesize{}$^{56}{\rm Ni}_{{\rm B}}$} & {\scriptsize{}{}${\rm E_{K}}$ }  & {\scriptsize{}{}${\rm (C/O)_{B}}$ }  & {\scriptsize{}{}${\rm IME_{B}}$ }  & {\scriptsize{}{}${\rm IGE_{B}}$ }  & {\scriptsize{}{}${\rm (C/O)_{U}}$ }  & {\scriptsize{}{}${\rm IME_{U}}$ }  & {\scriptsize{}{}${\rm IGE_{U}}$ }  & {\scriptsize{}{}$M_{{\rm Tot-U}}$}\tabularnewline
\cline{2-13} 
 & {\tiny{}{}$[{\rm M_{\odot}}]$  } & {\tiny{}{}{[}erg{]}  } & {\tiny{}{}{[}erg{]}  } & {\tiny{}$[{\rm M_{\odot}}]$  } & {\tiny{}{}{[}erg{]}  } & {\tiny{}{}$[{\rm M_{\odot}}]$  } & {\tiny{}{}$[{\rm 10^{-3}M_{\odot}}]$  } & {\tiny{}{}$[{\rm 10^{-3}M_{\odot}}]$  } & {\tiny{}{}$[{\rm 10^{-3}M_{\odot}}]$  } & {\tiny{}{}$[{\rm 10^{-3}M_{\odot}}]$  } & {\tiny{}{}$[{\rm 10^{-3}M_{\odot}}]$  } & {\tiny{}{}$[{\rm 10^{-3}M_{\odot}}]$  }\tabularnewline
\hline 
{\scriptsize{}A} & {\scriptsize{}{}0.53  } & {\scriptsize{}{}$4.8\times10^{46}$  } & {\scriptsize{}{}$3.2\times10^{47}$  } & {\scriptsize{}$3.1\times10^{-3}$} & {\scriptsize{}{}$3.5\times10^{48}$  } & {\scriptsize{}{}$0.49$  } & {\scriptsize{}{}$11$  } & {\scriptsize{}{}$5.7$  } & {\scriptsize{}{}$11$ } & {\scriptsize{}{}$9.0$  } & {\scriptsize{}{}$4.0$ } & {\scriptsize{}{}$24$}\tabularnewline
\hline 
{\scriptsize{}B} & {\scriptsize{}{}0.5  } & {\scriptsize{}{}$1.5\times10^{46}$  } & {\scriptsize{}{}$9.7\times10^{46}$  } & {\scriptsize{}$4.52\times10^{-3}$} & {\scriptsize{}{}$6.1\times10^{48}$  } & {\scriptsize{}{}$0.42$  } & {\scriptsize{}{}$36$  } & {\scriptsize{}{}$0.3$  } & {\scriptsize{}{}$34$  } & {\scriptsize{}{}$3.8$  } & {\scriptsize{}{}$6.1$  } & {\scriptsize{}{}$45$}\tabularnewline
\hline 
{\scriptsize{}C} & {\scriptsize{}{}0.55  } & {\scriptsize{}{}$8.2\times10^{45}$  } & {\scriptsize{}{}$2.6\times10^{46}$  } & {\scriptsize{}$5.25\times10^{-3}$} & {\scriptsize{}{}$2.8\times10^{48}$  } & {\scriptsize{}{}$0.49$  } & {\scriptsize{}{}$40$  } & {\scriptsize{}{}$0.44$  } & {\scriptsize{}{}$9.5$  } & {\scriptsize{}{}$0.5$  } & {\scriptsize{}{}$9.0$  } & {\scriptsize{}{}$19$}\tabularnewline
\hline 
{\scriptsize{}D} & {\scriptsize{}{}0.62  } & {\scriptsize{}{}$9.4\times10^{45}$  } & {\scriptsize{}{}$8.1\times10^{47}$  } & {\scriptsize{}$6.16\times10^{-3}$} & {\scriptsize{}{}$3.6\times10^{49}$  } & {\scriptsize{}{}$0.46$  } & {\scriptsize{}{}$28$  } & {\scriptsize{}{}$8.5$  } & {\scriptsize{}{}$120$  } & {\scriptsize{}{}$85$  } & {\scriptsize{}{}$9.6$  } & {\scriptsize{}{}$210$}\tabularnewline
\hline 
{\scriptsize{}E} & {\scriptsize{}{}0.62  } & {\scriptsize{}{}$2.3\times10^{46}$  } & {\scriptsize{}{}$6.9\times10^{47}$  } & {\scriptsize{}$2.84\times10^{-3}$} & {\scriptsize{}{}$1.4\times10^{49}$  } & {\scriptsize{}{}$0.48$  } & {\scriptsize{}{}$57$  } & {\scriptsize{}{}$24$  } & {\scriptsize{}{}$30$  } & {\scriptsize{}{}$23$  } & {\scriptsize{}{}$6.1$  } & {\scriptsize{}{}$90$}\tabularnewline
\hline 
{\scriptsize{}F} & {\scriptsize{}{}0.62  } & {\scriptsize{}{}$7.7\times10^{45}$  } & {\scriptsize{}{}$2.4\times10^{46}$  } & {\scriptsize{}$3.38\times10^{-3}$} & {\scriptsize{}{}$1.5\times10^{48}$  } & {\scriptsize{}{}$0.52$  } & {\scriptsize{}{}$76$  } & {\scriptsize{}{}$17$  } & {\scriptsize{}{}$4.8$  } & {\scriptsize{}{}$0.1$  } & {\scriptsize{}{}$3.3$  } & {\scriptsize{}{}$8.1$}\tabularnewline
\hline 
{\scriptsize{}G} & {\scriptsize{}{}0.73  } & {\scriptsize{}{}$1.8\times10^{46}$  } & {\scriptsize{}{}$8.8\times10^{46}$  } & {\scriptsize{}$2.68\times10^{-3}$} & {\scriptsize{}{}$1.2\times10^{49}$  } & {\scriptsize{}{}$0.65$  } & {\scriptsize{}{}$19$  } & {\scriptsize{}{}$9.1$  } & {\scriptsize{}{}$25$  } & {\scriptsize{}{}$23$  } & {\scriptsize{}{}$3.8$  } & {\scriptsize{}{}$52$}\tabularnewline
\hline 
{\scriptsize{}H} & {\scriptsize{}{}0.73  } & {\scriptsize{}{}$1.1\times10^{45}$  } & {\scriptsize{}{}$2.8\times10^{45}$  } & {\scriptsize{}$2.91\times10^{-3}$} & {\scriptsize{}{}$9.1\times10^{47}$  } & {\scriptsize{}{}$0.69$  } & {\scriptsize{}{}$32$  } & {\scriptsize{}{}$1.8$  } & {\scriptsize{}{}$2.8$  } & {\scriptsize{}{}$0.3$  } & {\scriptsize{}{}$0.5$  } & {\scriptsize{}{}$3.6$}\tabularnewline
\hline 
{\scriptsize{}I} & {\scriptsize{}{}0.73  } & {\scriptsize{}{}$7.5\times10^{45}$  } & {\scriptsize{}{}$2.1\times10^{46}$  } & {\scriptsize{}$8.25\times10^{-4}$} & {\scriptsize{}{}$2.5\times10^{48}$  } & {\scriptsize{}{}$0.64$  } & {\scriptsize{}{}$66$  } & {\scriptsize{}{}$8.9$  } & {\scriptsize{}{}$2.9$  } & {\scriptsize{}{}$5.5$  } & {\scriptsize{}{}$1$  } & {\scriptsize{}{}$9.4$}\tabularnewline
\hline 
{\scriptsize{}J} & {\scriptsize{}{}0.8  } & {\scriptsize{}{}$6.2\times10^{46}$  } & {\scriptsize{}{}$5.9\times10^{47}$  } & {\scriptsize{}$1.78\times10^{-3}$} & {\scriptsize{}{}$3.2\times10^{49}$  } & {\scriptsize{}{}$0.60$  } & {\scriptsize{}{}$82$  } & {\scriptsize{}{}$24$ } & {\scriptsize{}{}$40$  } & {\scriptsize{}{}$7.0$ } & {\scriptsize{}{}$45$  } & {\scriptsize{}{}$94$}\tabularnewline
\hline 
\end{tabular}
\end{table*}

\section{Discussion}

\subsection{Energetics and composition}

As described above the energetics and outflows in the NS-WD merger
we modeled are dominated by accretion energy and not by the nuclear
processes involved. Therefore the most energetic cases that are also
expected to produce the largest ejected masses are those in which
the most gravitational energy is released. This leads to several trends:
generally, the more massive the WD debris and the NS are, the more
energetic are the outflows and the thermonuclear energetics. Note, however, that the inner radius of
the debris disk also plays an important role. Immediately following the WD disruption the system is not in a steady accretion state, and the inner regions are initially empty. The position of the inner initial region of the debris disk is chosen as to coincide with the appropriate tidal radius given the NS and WD properties. However, the exact structure of the debris disk is more complex than our simplified models and we therefore checked the dependence of slightly by moving the disk position inwards or outwards (see ${\rm R_{0}/r_{t}}$ in Table \ref{tab:results}). As we show, the results depend on the chosen inner positions, but the overall behaviour is robust and is not strongly dependent on the exact choice. Closer-in debris disks
are more effective in allowing for material to accrete into close-in
orbits before outflows become efficient enough to quench further accretion.
The overall ejected mass and total energetic are therefore determined
by both the overall NS and WD mass and the inner radius of the disk;
these are well reflected in Table \ref{tab:results}.

Overall the the \emph{nuclear} energy deposited increases with increasing
density and with Helium content. These provide more favorable condition
for nuclear burning. The density is determined by the total mass of
the debris disk (the disrupted WD), the inner radius ($r_{d}$), the
disk scale height and to some extent the disk composition which affects
the EOS (see Eq. \ref{eq:rhodisk}). As can be seen in Table \ref{tab:results},
the nuclear energy produced follows these various trends; it increases
with increasing WD mass and Helium content, as well as with decreasing
disk scale height and inner radius. The tidal radius, which determines
the inner radius, depends on both the NS mass and the WD mass, as
well as on the WD composition, and hence more massive NSs (giving
rise to larger tidal radii) correspond to lesser production of nuclear
energy. The pure Helium WD case not shown) is the only case where
no nuclear burning is observed. This may appear counter-intuitive
given the lower temperatures and densities required to ignite Helium,
however such WDs have a lower mass and are much puffier, hence the
effective density of the debris disk they produce is far lower than
that of CO WD debris disk, explaining the result.

As discussed above, our results suggest that only a small fraction
of the WD debris disk is effectively ejected during the modeled evolution,
while most of the material expands to produce a more isotropic configuration,
but stays bound to the NS. The long-term evolution of such expanded
debris ``atmosphere'' is beyond the scope (and computational capabilities)
of the current study focusing on the immediate outcomes of the merger
following the disruption of the WD.

As already noted, nuclear burning is not the main driver of the outflows
in NS-WD mergers, and only a small fraction of the WD debris disk
experiences any nuclear burning. Hence the majority of the ejected
unbound material (identified through the comparison of the kinetic and potential energy of each mass element) shows the same composition as the original WD, i.e.
C/O composition (and He in the hybrid WD cases). However, most of
the unbound material is ejected through outflows from the inner regions,
where most of the nuclear burning occurs, and therefore, although
most of the ejected material is composed of C/O, it still contains
a significant, sometimes comparable fractions of burned material.
The latter is typically composed of both intermediate and Iron group
elements at comparable levels (Table \ref{tab:results}; see the composition table in the supplementary information for the more detailed composition).
Comparison of the results from the 19-isotope network and the extended 125-isotope network show very small, negligible differences when comparing the abundances of the same isotopes.

\subsection{Observational properties}

\subsubsection{Gamma-ray burst}

The total mass accreted on the NS in our simulations appears to be
small, as was also found by Metzger and collaborators \citep{2013ApJ...763..108F,2016MNRAS.461.1154M},
suggesting that such mergers are not likely to produce regular GRBs,
though one can not exclude ultra-faint long-GRBs extending for timescale
comparable to the accretion time scale of typically a few seconds.

\subsubsection{Optical transients}

Our overall results give rise to comparable ranges of ejected masses,
velocities and production of $^{56}{\rm Ni}$ as found in the simplified
1D model of \citet{Mar+16}. Consequently the observable expectations
are the same, namely the production of fast evolving, very faint transients,
which should peak at typical timescales of 6-7 days, and peak bolometric
luminosity of $10^{40}-10^{41}$ erg s$^{-1}$. Note, however, that
our multi-dimensional findings allow us to consider the overall structure
of the ejecta. We find that the structure of the outflows has a highly non-spherical
configuration, with most of the material ejected at intermediate inclinations,
and little mass ejected from the poles in a jet-like configuration
with a few times higher velocity than the typical ejecta. The latter
``jet''-ejecta contains a higher $^{56}{\rm Ni}$ fraction per unit
mass compared with the rest of the ejecta. The observational features
of such NS-WD mergers could therefore significantly vary as a function
of the viewing angle. 

Given their expected rapid evolution, such transients could be related
to the class(es) of fast evolving SNe \citep{deV+85,2010Sci...327...58P,Kas+10,2011ApJ...730...89P,Dro+13,Dro+14},
however, they are likely to be too faint to explain the the majority
of the luminous fast evolving SNe observed; if at all they are more
likely to be related to the faint end of such SNe such as SN 2010X.
Nevertheless, they might be too faint to even explain 2010X-like SNe,
and both their Helium and Aluminum content is small and may not explain
the He and Al lines identified in that SN \citep{Kas+10}. Transients
from NS-WD mergers may therefore present a completely different class
of SNe, that might be observed mostly in close-by galaxies using large
telescopes, possibly with next-generation surveys such as LSST.

As we show in \citet{Too+18b} the rates of NS-WD mergers could be
comparable to the inferred rate of fast evolving SNe \citep{Dro+14},
and the delay-time distribution show them to peak at early times of
hundreds of Myrs up to a Gyr, suggesting their typical host galaxies
to be late-type disk galaxies. We find that only a small fraction
of the mergers involve hybrid-WDs \citep{Zen+18} that may give rise
to some He ejection as observed in 2005E-like Ca-rich gap transients
\citep{2010Natur.465..322P}. These low rates and delay-time distributions
peaking at early times together with the extremely low-luminosity
and very little Ca production we find, likely exclude NS-WD mergers
as progenitors of the type Ib 2005E-like Ca-rich gap transients \citep{2010Natur.465..322P}.
The latter show clear evidence for He, are Ca-rich, explode at high
rates \citep{2010Natur.465..322P,Fro+18} and mostly explode at early-type
galaxies and old environments \citep{2010Natur.465..322P,Kas+12,Lym+13}
inconsistent with our findings for NS-WD mergers.

\subsubsection{Comparison with previous work}

The detailed studies most comparable with our work are those by \citet{2013ApJ...763..108F}
and \citet{Mar+16}. The former used a short-term 2D FLASH simulations
similar to ours, but used a simplified EOS and a single nuclear reaction.
They also did not consider the self-gravity of the disk, and did not
make a detailed post-analysis of the detailed composition. These issues
required them to use ad-hoc assumptions regarding detonations that
could not be resolved from the simulations. They also could not provide
a detailed composition analysis as done here. The latter work by \citet{Mar+16}
explore a simplified 1D model for the WD-debris disk. Such model required
them to use various assumptions regarding the wind-ejection which
can not be resolved in such models, and could not explore multi-dimensional
structure of the ejecta, however, it allowed them to provide detailed
models for the composition structure of the disk, and follow its evolution
up to late times.

{\bf
\begin{figure} 
\includegraphics[scale=0.3]{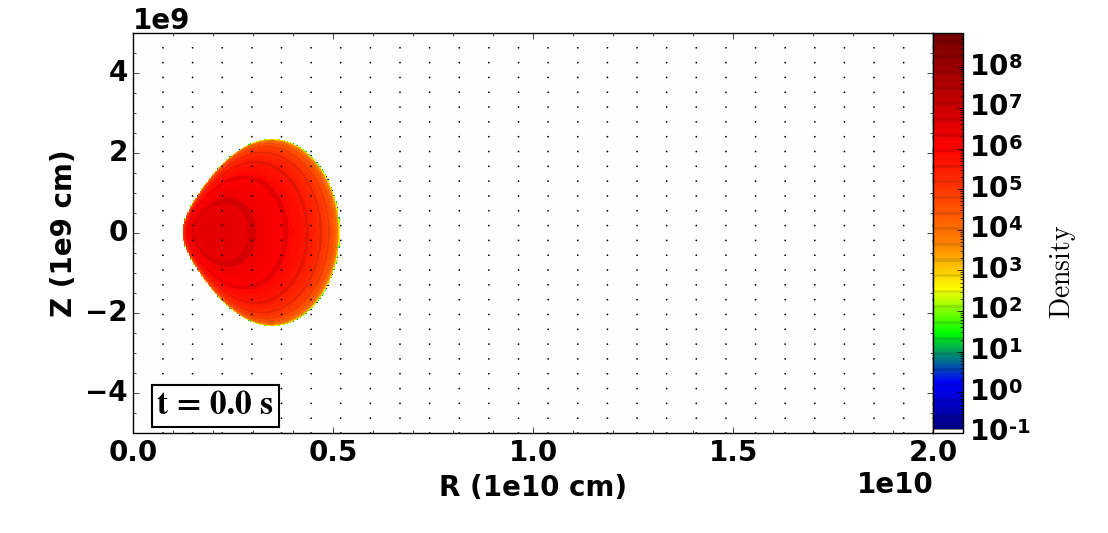}

\includegraphics[scale=0.3]{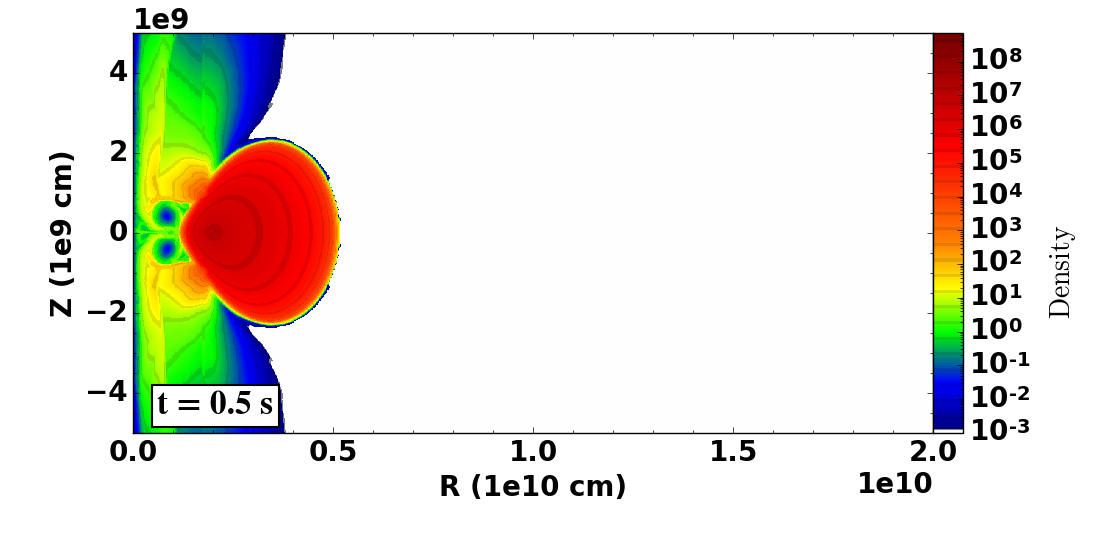}

\includegraphics[scale=0.3]{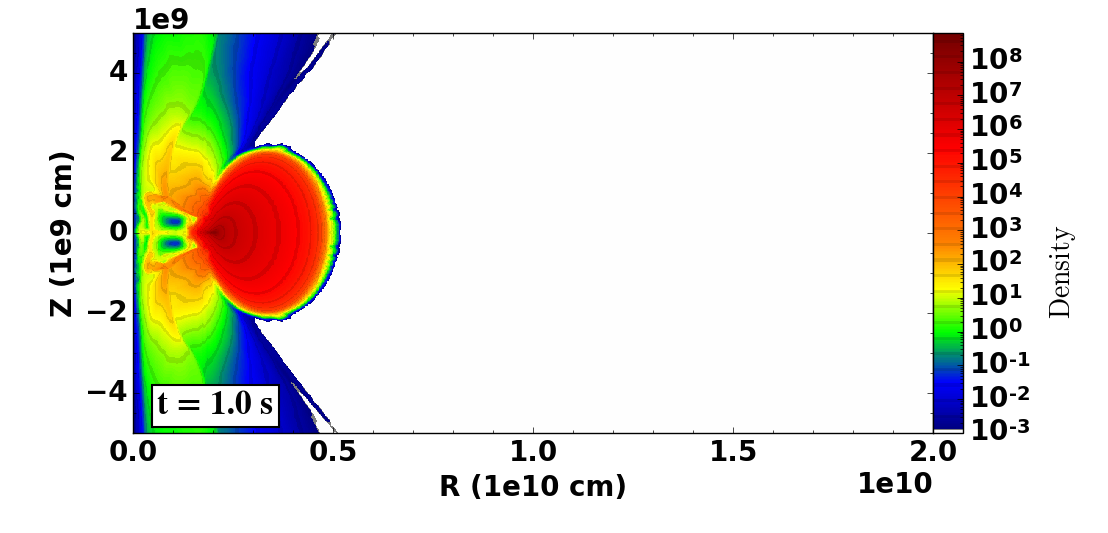}

\includegraphics[scale=0.3]{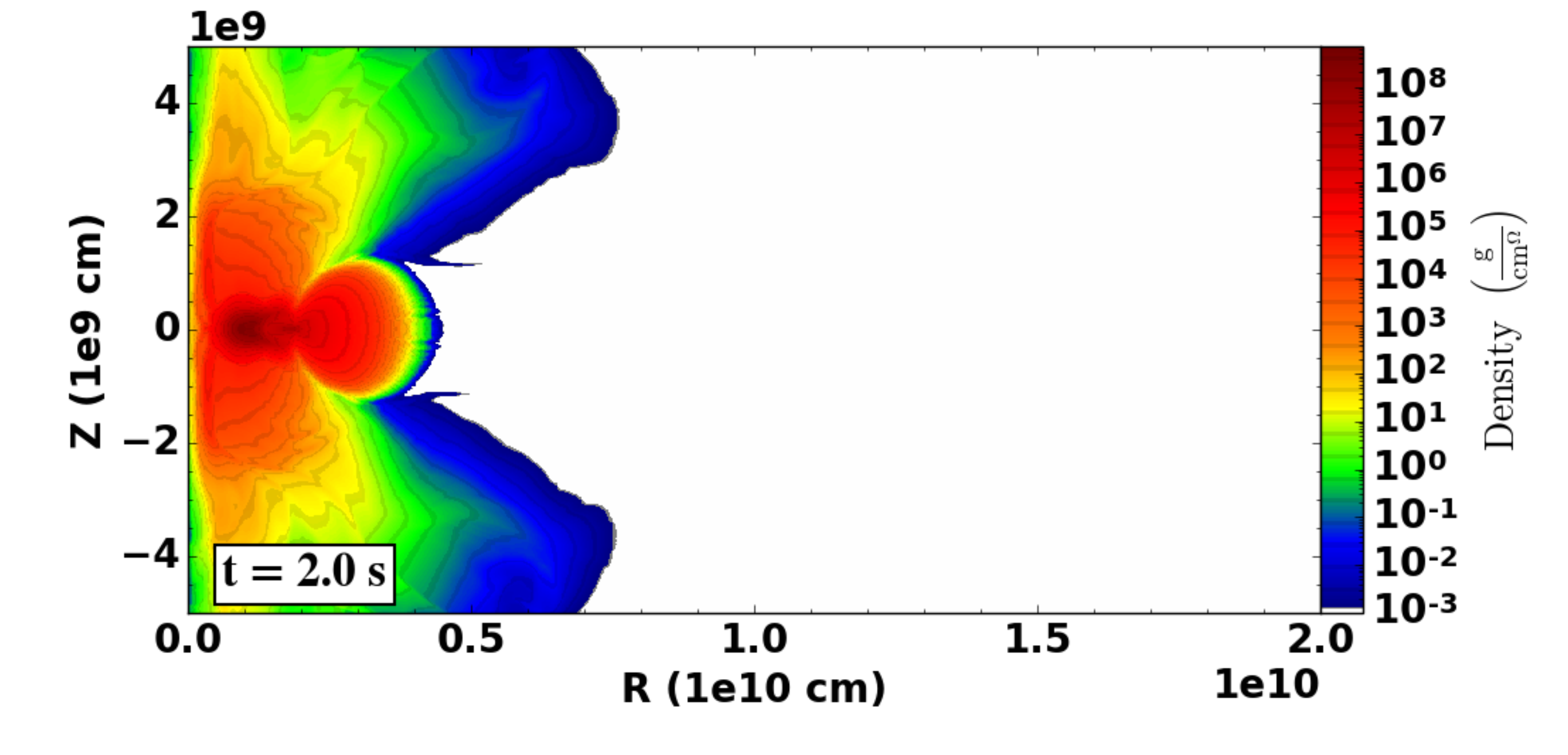}

\includegraphics[scale=0.3]{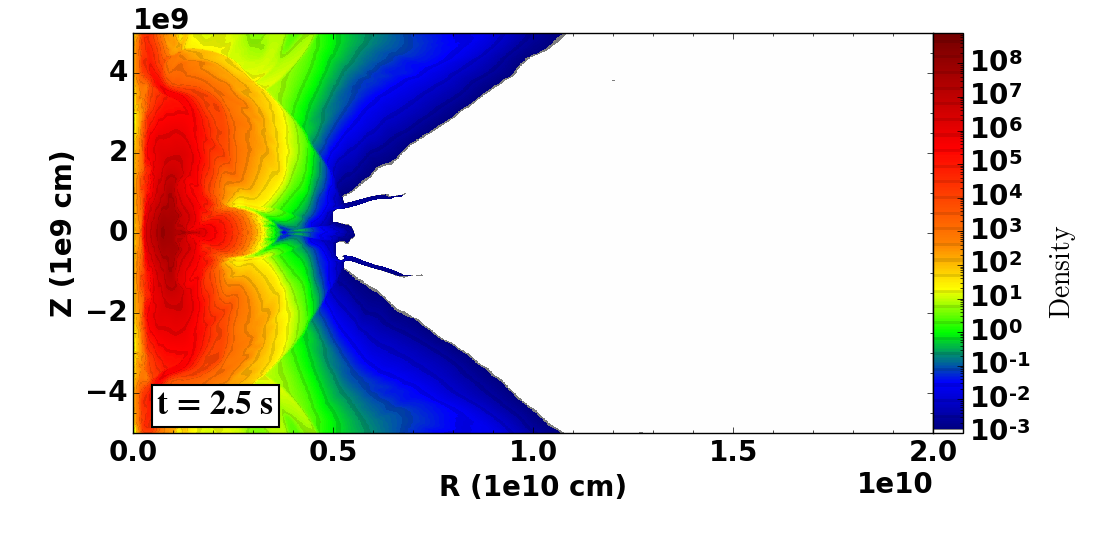}

\caption{The evolution of the WD debris at early times for a similar model as studied and shown in Fig. 1 of \citet{2013ApJ...763..108F}, but including self-gravity. As the disk evolves its inner parts collpase, producing a ``pinched''-like structure, not seen in the simulations without the self-gravity of the disk. A similar type of evolution is seen in all of our simulations.}
\label{fig:disk-self-grav}
\end{figure}}
Our models are therefore complementary to both these previous works
and extend them in several aspects. Our detailed 19-elements nuclear
network resolves the issues of nuclear energy and robustly show that
weak nuclear detonations are indeed produced, supporting the previous
models, albeit with somewhat lower energetics than envisioned by \citet{2013ApJ...763..108F}.
We also find that the self-gravity of the disk, not included before,
can significantly affect its evolution {\bf (see fig. \ref{fig:disk-self-grav})}, and the overall effects of
the detailed nuclear network, detailed EOS and self gravity give rise
to somewhat smaller energy contribution from the thermonuclear burning than suggested by the early
study of \citet{2013ApJ...763..108F}. Our long-term, lower resolution
models bridge the gap between the short-term 2D models at early times
explored by \citet{2013ApJ...763..108F} and the 1D long-term models
by \citet{Mar+16}, by allowing detailed 2D simulations extending
to late times. These results and our post-analysis detailed 125 elements
composition studies allow for comparison with the \citet{Mar+16}
study. Overall we find a good qualitative and quantitative consistency
between the overall composition results. Our models could therefore also be used as self-consistent
models to calibrate the hitherto assumed properties of winds in 1D
models. Moreover, our models also include details on the structure
of the wind and its jet-like polar configuration. 

Note that the direct comparison with \citet{Mar+16} is more difficult, given the different dimensionality and the very different assumptions taken. It is therefore difficult to say whether we should have expected exactly the same behaviour. In particular, we see significantly less accretion into the inner regions than found by them, though we should note that the typical grid boundary in our simulations is in the range of 1-2$\times10^{8}$, i.e. we do no resolve the innermost regions at the level possible in 1D models. Our simulations, however, do show significant nuclear burning and outflows at much larger scales than found in Margalit \& Metzger. This might not be surprising, since our simulations self-consistently resolve both the radial and vertical structure, unlike the 1D models, and the evolution could differ. In particular, following the inner collapse of the disk (not modeled in the 1D simulations, nor captured by the simulation of Fernandez \& Metzger 2013, which did not include the self gravity of the disk), the densities and temperatures become very high already at much larger scales than our resolves 10$^{8}$ cm, and are sufficiently high for the initiation of nuclear burning.
 
 In summary,  our models explored various regimes and initial conditions not explored before, enabling us to study the dependence of the merger
outcomes on the WD composition and mass, as well as on the detailed structure of of the WD debris disk. Though our models can not resolve the innermost regions close to the NS, as can be done in 1D simulations, they more self-consistently resolve the disk evolution and the wind mass-loss without introducing assumptions on the vertical evolution of the disk, nor on the wind mass-loss as in the 1D simulation. Our result suggest that nuclear burning initiates already at larger scales, and that the accretion to the inner regions is significantly smaller than inferred from the 1D simulations.

\section{Summary}

In this study we explored the early-time evolution of NS-WD mergers
using a 2D hydrodynamical simulation of the debris disk of a disrupted
WD around a NS. We made use of an extended nuclear-network to follow
nuclear burning in such models, and explored a wide range of initial
conditions and combinations of masses and compositions of NSs and
WDs. We find that such mergers are mostly driven by the gravitational
energy released in the accretion process, giving rise to mass ejection
through strong winds launched from the inner regions of the accretion
disk. These produce a ``jet''-like configuration of fast polar-winds
which eject little-mass, and somewhat slower winds at intermediate
inclinations which carry most of the ejected material. We find support
for earlier claims that weak nuclear detonations are produced in the
inner regions of the accretion disk, thereby producing intermediate
and Iron-group elements. Nevertheless, the radioactive $^{56}{\rm Ni}$
production is limited to the range of a few $10^{-4}-10^{-3}$ ${\rm M_{\odot}}$
and could only give rise to very faint transients, with the nuclear
energy providing only $\sim1-10\%$ of the total kinetic of the ejected
material. The properties of such transients suggest that to be a different
class than any of the observed SNe, though they might be related to
the faint tail of the observed classes of fast-evolving SNe. Their
overall properties are inconsistent with those of 2005E-like type
Ib faint Ca-rich SNe, and likely exclude them as progenitors of such
SNe. 

\section*{Acknowledgements}
We acknowledge support from the Israel science foundation I-CORE
1829/12 grant and the Rafael Science foundation. ST acknowledges support
from the Netherlands Research Council NWO (grant VENI {[}\#639.041.645{]}).
We thank Brian Metzger, Oded Papish, Noam Soker, Ben Margelit and
Sagiv Shiber for stimulating discussions.






\bibliographystyle{mnras}


\appendix
\section{Elements table}
See elements table in the supplementary material.
\bsp	
\label{lastpage}
\end{document}